\documentclass[12pt,preprint]{aastex}

\slugcomment{Astrophysical Journal}

\shorttitle{Composition of Primary Cosmic-Ray Nuclei at High Energies}
\shortauthors{Ave et al.}

\begin{document}


\title{Composition of Primary Cosmic-Ray Nuclei at High Energies}


\author{M. Ave, P.J. Boyle, F. Gahbauer\altaffilmark{a},
C. H\"{o}ppner\altaffilmark{b}, J.R. H\"{o}randel\altaffilmark{c},
M. Ichimura\altaffilmark{d}, D. M\"{u}ller, A. Romero
Wolf\altaffilmark{e}}

\affil{Enrico Fermi Institute, The University of Chicago, \\933 E 56th
Street, Chicago, IL 60637, USA}


\altaffiltext{a}{Currently at: University of Latvia, Latvia.}
\altaffiltext{b}{Currently at: Technische Universit\"{a}t M\"{u}nchen, Germany.}
\altaffiltext{c}{Currently at: Radboud University Nijmegen, The Netherlands.}
\altaffiltext{d}{Currently at: Hirosaki University, Japan.}
\altaffiltext{e}{Currently at: University of Hawaii, USA.}


\begin{abstract}
The TRACER instrument (``Transition Radiation Array for Cosmic
Energetic Radiation'') has been developed for direct measurements of
the heavier primary cosmic-ray nuclei at high energies. The instrument
had a successful long-duration balloon flight in Antarctica in
2003. The detector system and measurement process are described,
details of the data analysis are discussed, and the individual energy
spectra of the elements O, Ne, Mg, Si, S, Ar, Ca, and Fe (nuclear
charge Z=8 to 26) are presented. The large geometric factor of TRACER
and the use of a transition radiation detector make it possible to
determine the spectra up to energies in excess of 10$^{14}$ eV per
particle. A power-law fit to the individual energy spectra above 20
GeV per amu exhibits nearly the same spectral index ($\sim$ 2.65 $\pm$
0.05) for all elements, without noticeable dependence on the elemental
charge Z.

\end{abstract}




\section{Introduction}
\label{sec:intro}

The energies of cosmic rays observed near earth extend over a very
wide range, from below 10$^8$ eV to more than 10$^{20}$ eV per
particle.  Up to about 10$^{10}$ eV per amu, the particle energies and
intensities are significantly affected by solar modulation, but for
the nearly ten remaining decades, the arriving particles are believed
to represent the ambient cosmic-ray population in the local galaxy.
Over this range, the total cosmic-ray intensity (the ``all particle''
differential energy spectrum, see, for instance, the compilation of
\citet{cronin97}) decreases monotonically by more than 30 orders of
magnitude, initially as a power law $\propto$ E$^{-2.7}$, steepening
slightly to E$^{-3.0}$ at the ``knee'' above 10$^{15}$ eV per
particle, and exhibiting small changes in slope again (a ``second
knee'' and an ``ankle'') in the 10$^{18}$ -- 10$^{19}$ eV region.
Besides these relatively minor changes, the overall spectrum is
remarkably featureless, even though a variety of processes may
contribute to the cosmic-ray flux. The general consensus is that the
particles below the knee are generated inside our Galaxy and are
contained by the galactic magnetic fields for millions of years, while
at the highest energies ($\geq$ 10$^{18}$ eV) the gyration radii
become so large that galactic containment is no longer effective, and
that the cosmic rays may then be of extragalactic origin.

The all-particle energy spectrum refers to the overall spectrum of
cosmic rays without differentiating the individual components.  To
obtain deeper insight, details of the composition of the cosmic-ray
particles must be studied.  For instance, very accurate measurements
of not only the elemental, but also the isotopic composition are now
available at low energies ($\leq$ 10$^{9}$ -- 10$^{10}$ eV per
particle).  These have determined such important quantities as the
average containment time of cosmic rays in the galaxy from
observations of radioactive clock-nuclei ($\tau \approx 1.5 \times
10^{7}$ y at these energies (first reported by \citet{gmunoz75}; more
recent work by \citet{yanasak01}), and they have excluded fresh
supernova ejecta as source material of cosmic rays from the abundances
of Co- and Ni-isotopes \citep{wiedenbeck99}.  At relativistic energies
only the abundances of the elemental species, without isotopic detail,
have been accessible to measurements (e.g. \citet{engelmann90}), but
even these encounter increasingly severe systematic and statistical
uncertainties as the energy increases.  Very few direct observations
(e.g. \citet{muller91}) have provided composition detail with
single-element resolution above ~10$^{13}$ eV per particle.  The work
described in this paper represents an attempt to improve this
situation, and to determine the energy spectra of individual elemental
species up to much higher energies.

The current paradigm of the origin of galactic cosmic rays postulates
first-order Fermi acceleration in interstellar shock fronts from
supernova (SN) explosions as the most potent contributor to the
cosmic-ray flux below the knee. This mechanism, first proposed by
\citet{bell78}, must be very efficient indeed, since as much as $\sim$10\%
of the total kinetic energy released in supernova explosions is
required to sustain the cosmic-ray flux. The SN shock acceleration
mechanism predicts a source energy spectrum in the form of a power law
$\propto$ $E^{-\Gamma}$, with a spectral index $\Gamma \approx 2.0$
for strong shocks.  Such a spectrum is much harder than the $E^{-2.7}$
spectrum that seems to be typical for most primary cosmic-ray nuclei
below the knee.  This difference in spectral slope can be explained if
the propagation and containment of cosmic rays in the galaxy are
energy-dependent.  Such a behavior had indeed been inferred, even
before the SN shock acceleration model was formulated, on the basis of
measurements of the ``L/M ratio'', i.e., of the abundance of
secondary, spallation-produced cosmic rays (such as the light elements
Li, Be, B) relative to their primary parents (such as C and O),
\citep{juliusson72}. One concludes from such measurements that the
average amount of interstellar matter encountered by cosmic-ray nuclei
during their propagation through the galaxy decreases with energy. The
effect is often parameterized by a propagation pathlength $\Lambda$,
which depends on energy as $\Lambda \propto E^{-0.6}$ for relativistic
nuclei \citep{engelmann90}. The primary cosmic-ray spectrum at the
source would then be approximately $\propto E^{-2.1}$ or $E^{-2.2}$,
close to what the shock acceleration model predicts.  This fact
provides strong but indirect evidence for the validity of the shock
acceleration model.

However, it must be kept in mind that current measurements of the L/M
ratio do not extend beyond 10$^{11}$ eV per amu (or about 10$^{12}$ eV
per particle).  Hence, there is an extended region of energies
(10$^{12}$ -- 10$^{15}$ eV per particle) where the SN shock
acceleration model is assumed to be valid, albeit without the benefit
of much observational support.  The SN shock acceleration process is
expected to become inefficient at energies around $Z \times$ 10$^{14}$
eV (where $Z$ is the atomic number of a cosmic-ray nucleus)
\citep{lagage83}, but this limit is not observationally confirmed.
One might refer to the steepening of the all-particle spectrum at the
"knee" as evidence, but even then the origin of cosmic rays with
energies far beyond the knee remains a mystery.

The SN-shock acceleration model has recently found strong
observational support through the detection of TeV gamma rays from
shell-type SN-remnants \citep{aharonian04}, although there remains
a debate whether the parent particles of these gamma rays are electrons
or, indeed, protons and nuclei \citep{berezhko06}. 

Clearly, the present observational evidence about high-energy cosmic
rays is inadequate to provide answers to some of the most fundamental
questions about their origin and galactic propagation, and there
exists an undeniable need for improved measurements at higher
energies. The TRACER program has been developed to provide direct
measurements of the energy spectra of the heavier cosmic-ray nuclei
through balloon flights above the atmosphere, up to energies between
10$^{14}$ and 10$^{15}$ eV per particle [The acronym TRACER stands for
``Transition Radiation Array for Cosmic Energetic Radiation'']. Higher
energies are currently studied through indirect observations with
ground-based air-shower installations. However, attempts to obtain
composition details from air-shower observations are affected by a
number of systematic uncertainties.  One of the goals of TRACER is to
approach the energy region of air shower measurements, with the hope
of providing some cross-calibration with the indirect measurements.

\section{Observational Technique}
\label{sec:technical}

\subsection{General Principles}

A successful cosmic-ray measurement above the atmosphere must
determine at least two quantities for each particle: the charge $Z$ and
the energy $E$.  Measurement of $Z$ is commonly accomplished by utilizing
the fact that all electro-magnetic interactions scale with $Z^{2}$.
Hence, devices that measure the ionization energy loss (dE/dx) such as
plastic scintillators or gas proportional counters, or that measure
the intensity of Cherenkov light, are well suited and are used in
TRACER. 

The energy determination is a greater challenge, mainly because of the
large detector areas that are needed at high energy.  Extrapolations
from lower energies indicate that a successful measurement of nuclei
heavier than helium requires detectors with exposure factors of at
least several 100 m$^{2}$ sr days in order to approach the 10$^{15}$
eV per particle region.  Such exposures can be accomplished with
long-duration (i.e., several weeks) balloon flights if the detectors
have sensitive areas of a few square meters.  A classic calorimeter
that measures the total energy absorbed in matter after a particle
interacts would become much too massive if it were to have an area of
this magnitude. Hence, TRACER uses electromagnetic interactions not
just to determine $Z$, but also to measure the energy (or more
exactly, the Lorentz-factor $\gamma \approx E/mc^{2}$): measurement
of Cherenkov light produced in an acrylic Cherenkov counter at low
energies (up to a few 10$^{9}$ eV per amu), measurement of the
specific ionization in gases at intermediate energies (10$^{10}$ to
10$^{12}$ eV per amu), and measurement of transition radiation at the
highest energies ($>$ 5 $\times$ 10$^{11}$ eV per amu).

\begin{figure}[t]
\begin{center}
\includegraphics[width=0.98\textwidth]{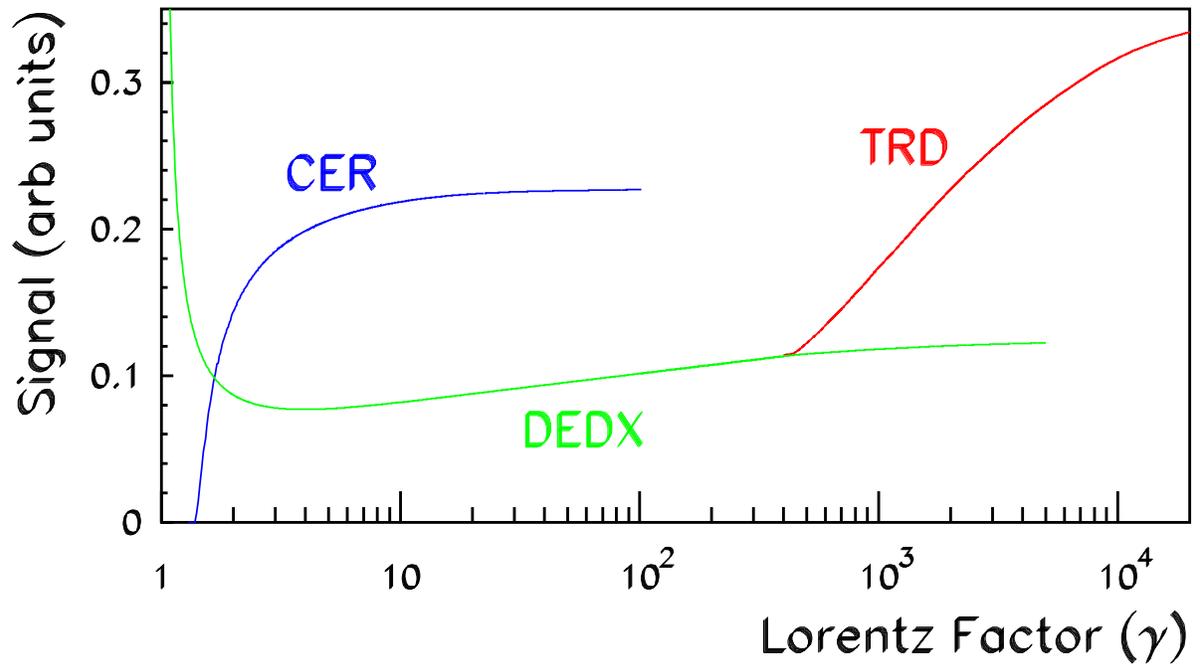}
\end{center}
\caption{Energy response of detection techniques used in TRACER:
Cherenkov counter (CER), energy loss in gases (DEDX), and Transition
Radiation (TRD).}
\label{resp}
\end{figure}

Figure \ref{resp} shows the typical energy responses for the detectors
used in TRACER.  The light intensity of an acrylic Cherenkov counter
(refractive index = 1.49) increases quickly with energy above a
threshold of 0.325 GeV per amu ($\gamma \approx$ 1.35) until it
approaches saturation above 20 GeV per amu ($\gamma \approx$ 22).  The
ionization loss in gas decreases with energy, reaches a minimum at
$\gamma$ = 3.96, and then increases logarithmically with $E$.
Transition radiation (TR) x-rays are generated in a multi-surface
radiator (mats of plastic fibers in TRACER) and are detected in a gas
proportional counter.  Hence, the TR detector measures an ionization
loss signal, and at sufficiently high energies ($>$ 5 $\times$
10$^{11}$ eV per amu), a superimposed signal due to the absorption of
TR photons.  The TR signal rises rapidly with the Lorentz factor
$\gamma$ and reaches saturation in the $\gamma$ = 10$^{4}$ -- 10$^{5}$
region.  The use of TR to measure the highest cosmic-ray energies was
first implemented for the CRN detector in 1985 \citep{lheureux90}, but
remains unconventional.  It permits the construction of detectors of
very large area.  As the TR intensity increases with $Z^{2}$, this
technique is well suited for measurements of the heavier cosmic rays
($Z >$ 3), but not applicable to protons or He-nuclei where the
signals are affected by large statistical fluctuations.

Note that the response curves shown in figure \ref{resp} for the
detectors of specific ionization and TR are double-valued: large
signals can either be produced by high-energy particles, or by
particles below minimum ionization. The TRACER instrument must be
launched at latitudes with low geomagnetic cutoff energies where a
significant cosmic-ray intensity arrives with sub-relativistic
energies. The inclusion of a plastic Cherenkov counter which
identifies the low-energy particles is then essential in order to
remove the degeneracy of ionization and TR response.  Overall, the
counter combination chosen for TRACER measures the energies of
cosmic-ray nuclei over a very wide range, in excess of four orders of
magnitude.

\subsection{The TRACER Instrument}

\begin{figure}[t]
\begin{center}
\includegraphics[width=.9\textwidth]{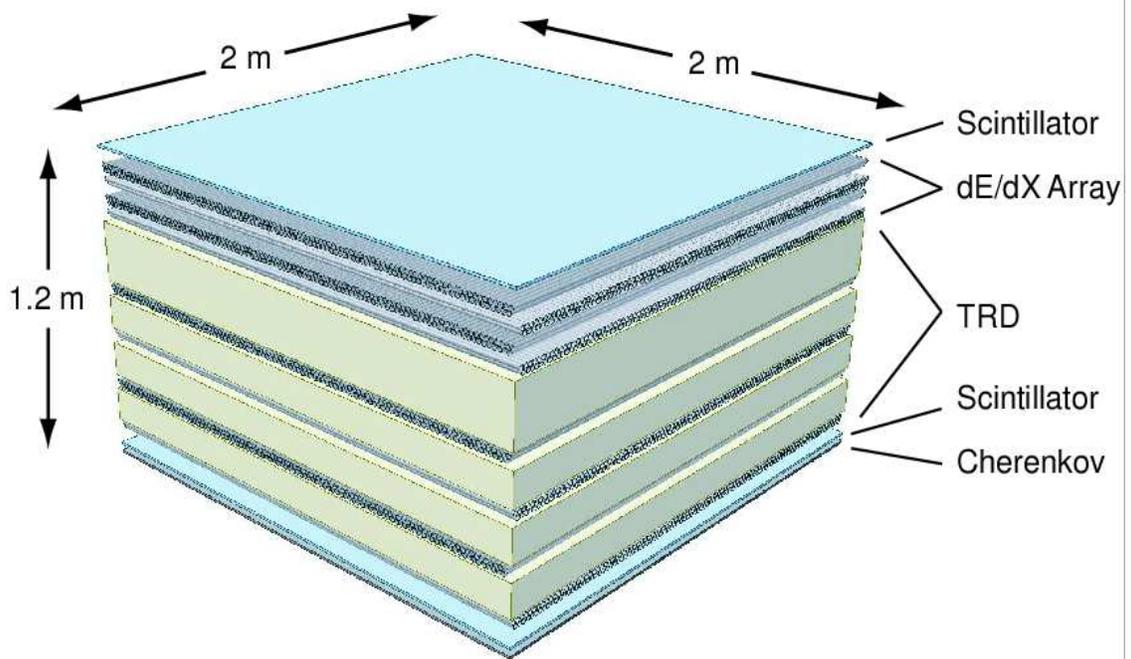}
\end{center}
\caption{Schematic drawing of TRACER}
\label{tracer}
\end{figure}

A schematic cross-section of the TRACER instrument is shown in figure
\ref{tracer}.  The detector elements are rather large in area (206 cm
$\times$ 206 cm), and are vertically separated by 120 cm.  This leads
to an overall geometric factor of about 5 m$^{2}$sr.  The instrument
has been previously described \citep{gahbauer04}, and a detailed
technical description of the detector elements is currently being
prepared for a separate publication.  Here, we will present a brief
summary of the relevant properties of the instrument.

As figure \ref{tracer} shows, the individual components of TRACER are,
from top to bottom:

1. A plastic scintillator sheet (BICRON-408 material), with active
   area 200 cm $\times$ 200 cm, and 0.5 cm thick\footnote[1]{Due to
   the interspersed wavelength shifter-bars, the geometric area of
   these counters is slightly larger than the active area.}.  The
   scintillator is viewed by 24 photomultiplier tubes (PMT) via
   wavelength shifter bars.

2. Four double layers of single-wire proportional tubes, oriented in
   two orthogonal directions.  Each tube is 200 cm long and 2 cm in
   diameter.  The tube walls are wound from aluminized mylar and are
   about 127 $\mu$m thick.

3. Another four double layers of proportional tubes of identical
   design, but each double layer located below blankets of plastic
   fiber material, which form a radiator to generate transition
   radiation.

   All tubes are filled with a mixture of xenon and methane (at equal
   parts by volume) and are operated in flight at a pressure of 0.5
   atmospheres.

4. A second scintillator of identical design, but read out by just 12
   PMTs, and located below the lowest layer of proportional tubes.

5. A Cherenkov counter, with active area 200 cm $\times$ 200 cm and
1.27 cm thick, at the bottom of the detector stack. The Cherenkov
material consists of an acrylic plastic (refractive index = 1.49),
doped with wavelength shifter material. Signals are read out via
wavelength-shifter bars connected to 24 PMTs (as for the scintillator
on top). 

The two scintillators serve as triggers for the instrument and,
together with the Cherenkov counter, determine the elemental charge
$Z$ of individual cosmic-ray nuclei traversing the instrument. The
proportional counter arrays measure the energy (or Lorentz-factor $
\approx E/mc^{2}$) of the cosmic-ray particles at high energies, and
they determine the trajectory of each particle traversing the
instrument in two projections.  The signals in the top four double
layers simply are a measure for the ionization energy loss ($dE/dx$)
of the particle. The small logarithmic increase above the
minimum-ionization level provides an estimate of the particle energy
up to Lorentz factors of a few hundred. The lower four double layers
of tubes form a transition radiation detector (TRD).  As the response
curves in figure \ref{resp} show, the TRD signals are identical to the
dE/dx signals of the tubes above, except at the highest energies
($\gamma$ $>$ 400), where the rapid increase of the signal with
increasing $\gamma$ permits an energy measurement. The signals of the
proportional tubes (a total of about 1600) are processed through VLSI
chips (``Amplex'', \citet{beuville90}).  The finite dynamic range of
these devices restricts the range of elemental species for which data
could be recorded in this flight, from oxygen ($Z$ = 8) to iron ($Z$ =
26).

The TRACER instrument derives its heritage from the CRN detector that
was developed by our laboratory in the 1980s for space flight
\citep{lheureux90}.  In fact, the radiator material used for the
TRACER-TRD is identical to that of CRN. However, CRN employed
thin-window multiwire proportional chambers to detect transition
radiation, and hence, needed a heavy pressurized container.  The
proportional tubes used in TRACER instead can easily withstand
external low pressure; and thus, a pressurized gondola is not required
and not used for TRACER.

\section{Balloon Flight and Instrument Performance}
\label{sec:inst}
\begin{figure}
\begin{center}
\includegraphics[width=.9\textwidth]{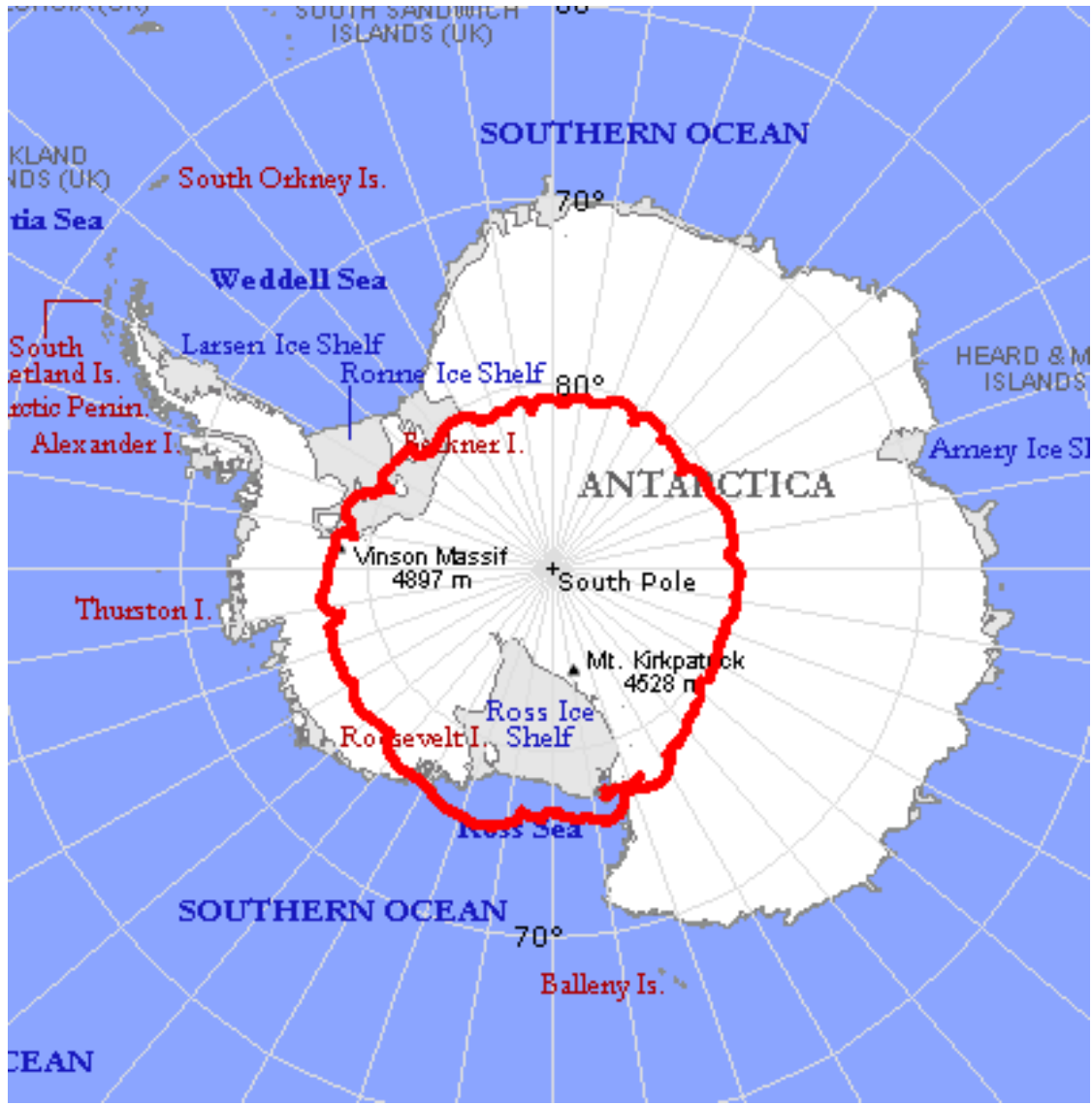}
\end{center}
\caption{Trajectory of 2003 TRACER Antarctic Long Duration Balloon Flight.}
\label{map}
\end{figure}

TRACER had first a one-day balloon flight in Fort Sumner, New Mexico,
in 1999.  The results from this flight have been published
\citep{gahbauer04}.  A circumpolar long-duration balloon (LDB) flight
was originally planned for a flight in the northern hemisphere with a
launch from Alaska, but could not be conducted because of the lack of
relevant international agreements.  Therefore, an LDB flight had to be
planned in Antarctica, but had to wait until 2003, when a new launch
vehicle commensurate with the weight of TRACER (6,000 lbs., including
ballast, telemetry, and balloon-related instrumentation) became
operational. TRACER was launched near McMurdo, Antarctica, on December
12, 2003.  During its two-week flight (see figure~\ref{map}) at an
average height of 37.75 km (3.9 g/cm$^2$) with excursions not
exceeding $\pm$ 1.75 km, it collected data from a total of 5 $\times$
10$^{7}$ cosmic-ray events.  The data were transmitted at 500 Mbit/sec
to the ground (when the instrument was within telemetry range) and
stored on six disks on-board and recovered after termination of the
flight.

Overall, the instrument performed very well during this flight. In
particular, the entirely passive thermal insulation kept the
temperatures of all detector components close to room temperature,
with diurnal variations of at most a few degrees C. There was no
indication of a deterioration of the performance of the proportional
tubes due to gas poisoning for the duration of the flight. This result
would be of great practical importance if flights of such devices for
much longer duration (for instance in space) are anticipated: there
seems to be no need for gas-purging at the time scales of at least a
few weeks, except for the provision of make-up gas to correct for
minor gas leaks. The data acquisition system operated very
efficiently, with a dead time of just 6\%, but the total time during
which data could be recorded was limited to ten days, due to a
catastrophic failure of the rechargeable lithium batteries (which were
used to buffer the solar-power system) towards the end of the first
orbit of the balloon around the South Pole.

\section{Data Analysis}
\label{sec:analysis}

\subsection{Trajectory Reconstruction}
\label{sec:trajectory}
The first step in the analysis of the data is the accurate
reconstruction of the trajectory of each cosmic-ray particle through
the instrument. The entire array of proportional tubes is used for
this purpose. It consists of four double layers of tubes in each the
x- and y- direction. Hence, ideally a trajectory is characterized by
eight tube ``hits'' in each projection. In practice, sometimes the
particle may pass between adjacent tubes of a given layer and not
generate a hit in this layer. In addition, spurious signals in tubes
outside the particle trajectory may be recorded.

As a first estimate, the path is obtained using the center of each of
the tubes hit in an event. All possible combinations of these tubes
are fit with straight lines in the $X$ and $Y$ projection, and the
combination in each projection with minimum $\chi^2$ per degree of
freedom and maximum number of tubes used is kept. This method allows
an accuracy of 5 mm in track position. The procedure is quite
efficient: less than 5~\% of cosmic-ray nuclei are missed in the
process.

A preliminary estimate of the charge of each particle is then
performed using the technique described in the following section. A
subset of the data from which we expect the same average signal in any
tube (high energy $O$ or $Ne$ nuclei) is then used to correct for
deviations of the tube positions from their ``ideal'' positions, and
to match signal gains within 5~\% or less. Corrections due to
variations in the tube gain with time either due to temperature
changes or small gas leakages are also calculated and applied. In
general, gain variations in a given tube are negligible \emph{along}
that tube. Finally, the tube signals are corrected to account for
minor non-linearities in the front end electronics at large
signals. All these corrections are necessary to reduce the systematic
uncertainties in the tube signals below the level of unavoidable
physical fluctuations (``Landau fluctuations'') in the energy
deposited in the gas.

As a final step, the trajectory is refined by using the fact that the
energy deposit in each tube is proportional to the track length in
that tube. Taking this fact into account, an accuracy of 2 mm in the
lateral track position is achieved, which corresponds to a 3~\%
uncertainty in the total pathlength through all the tubes. The
accuracy and efficiency of this method are verified with a GEANT4
simulation \citep{agostinelli03} of the instrument.

The accurate knowledge of the particle trajectories is essential to
the analysis for two reasons: first, it permits corrections of the
scintillator and Cherenkov signals due to spatial non-uniformities and
zenith-angle variations (see section \ref{sec:charge}). Second, it
makes possible to determine accurately the average energy deposit of a
cosmic-ray particle when it traverses the tubes. Specifically, the
average signals are as follows:

In the top four double layers, the average specific ionization is defined as 

\vspace{0.1cm}
\begin{center}
\begin{equation}
\label{dedxeq}
\langle \frac{dE}{dx} \rangle = \frac{\sum_{i=1}^{i=8} \Delta{E}_i}{\sum_{i=1}^{i=8} \Delta{x}_i}
\end{equation}
\end{center}

and in the bottom four double layers (which form the TRD), the average
signal is

\begin{center}
\begin{equation}
\label{dedxtreq}
\langle \frac{dE}{dx}+TR \rangle = \frac{\sum_{i=9}^{i=16} \Delta{E}_i}{\sum_{i=9}^{i=16} \Delta{x}_i}
\end{equation}
\end{center}
\vspace{0.1cm}

Here, $\Delta{E}_{i}$ and $\Delta{x}_{i}$ refer to energy deposit and
pathlength, respectively, in tube number $i$. The two quantities defined
in (\ref{dedxeq}) and (\ref{dedxtreq}) are, of course, identical
within fluctuations at energies where no transition radiation is
detectable ($TR = 0$). In order to exclude large fluctuations
associated with short pathlengths, we remove tubes with $\Delta{x}_{i}
<$ 1 cm from the summation.
 
\subsection{Charge Analysis}
\label{sec:charge}

\begin{figure}[ht]
\begin{center}
\includegraphics[width=.99\textwidth]{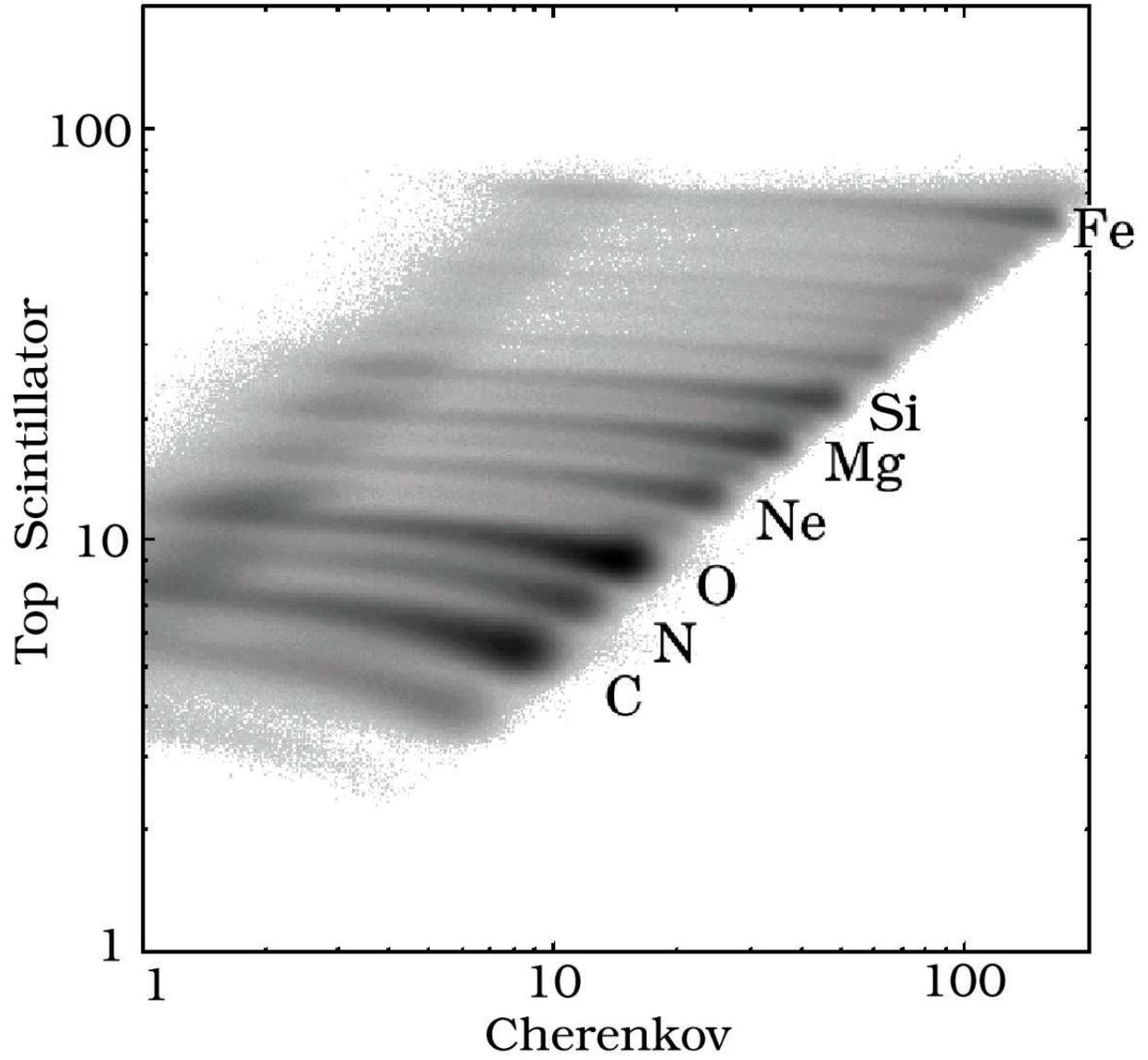}
\end{center}
\caption{Scatter plot of top scintillator vs. Cherenkov signals in arbitrary units.}
\label{CerScint}
\end{figure}

As a first step to determine the charge of each particle, the signal
recorded by each PMT is corrected to account for the spatial
non-uniformities in the counter responses. These corrections are
derived using the tracking information together with response maps
recorded with muons at ground and verified with the flight data
themselves in an iterative procedure. Subsequently, for each event the
average signal of all PMTs in each counter is calculated, and then
normalized to vertical incidence. Figure \ref{CerScint} shows a
scatter plot of the $top$ scintillator signal vs the Cherenkov
signal. It is apparent that particles with the same charge are
clustered along lines. The position along the line depends on the
primary energy: a concentration of events around maximum Cherenkov and
minimum scintillation signals corresponds to energies of minimum
ionizing particles and above, and the energies decrease for events
towards the left of this concentration.

\begin{figure}[ht]
\begin{center}
\includegraphics[width=.99\textwidth]{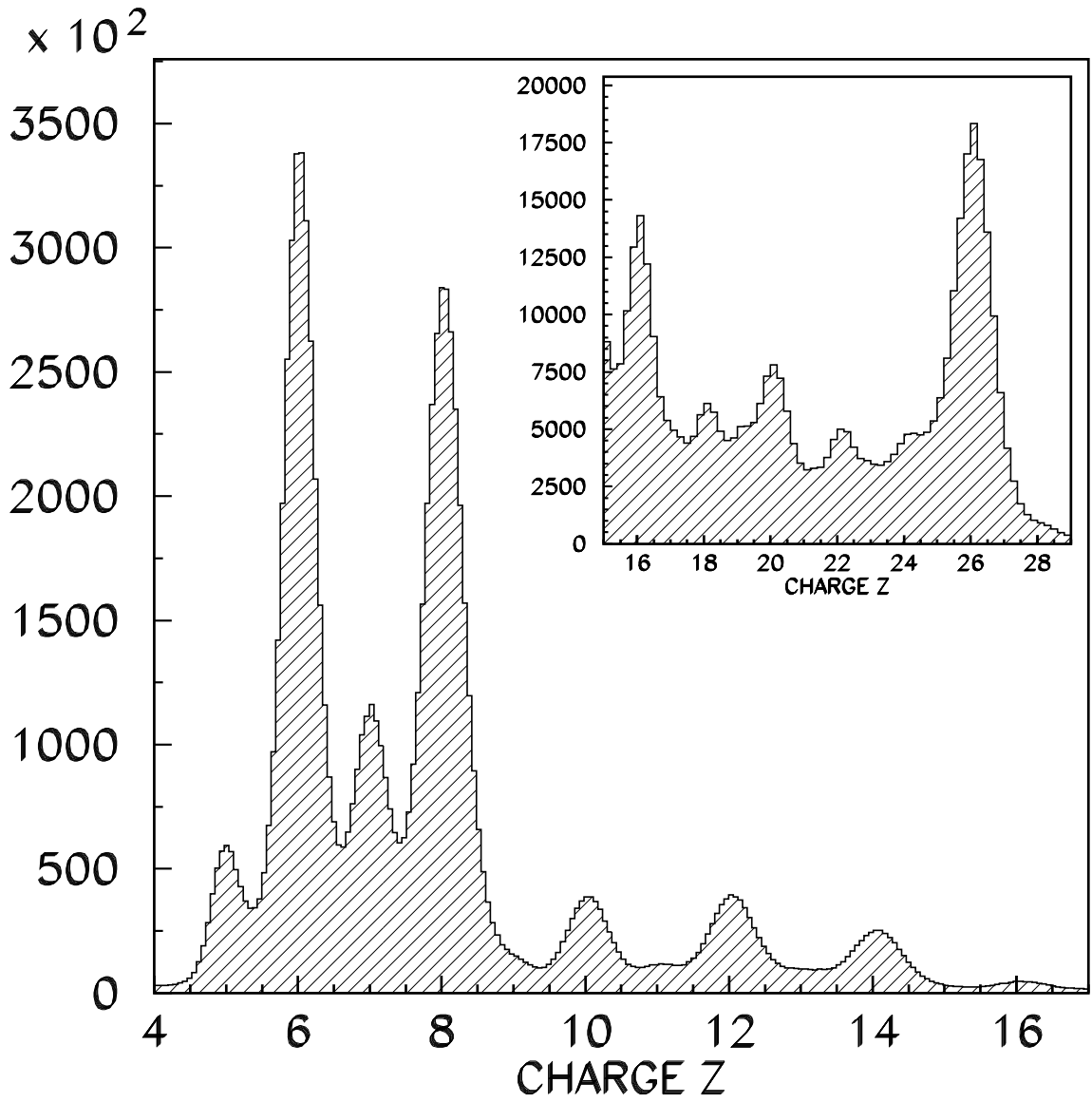}
\end{center}
\caption{Charge histogram for all events measured in flight.}
\label{Zhisto}
\end{figure}

The scintillator signals depend on particle charge as $Z^{1.65}$, due
to non-linear effects in the light yield, while the Cherenkov signals
follow the expected $Z^2$ dependence.  The residual scintillation in
the Cherenkov counter is visible in figure \ref{CerScint}, and is
subtracted out in the analysis. Figure \ref{Zhisto} shows a charge
histogram for all charges obtained from summing along lines of
constant charge in figure \ref{CerScint}. The charge resolution
evolves from 0.25 charge units for $Z$=8 ($O$ nuclei) to 0.5 charge
units for $Z$=26 ($Fe$ nuclei) at low energies ($<$ 10 GeV per amu),
and increases slightly to 0.3 and 0.6 charge units, respectively, at
the highest energies.

Note that figure \ref{Zhisto} is generated just on the basis of the
signals of the top scintillator, and thus depicts a histogram of
$Z_{top}$. One can, in the same fashion, construct a histogram of
$Z_{bot}$, using the signals of the bottom scintillator. For the
further analysis, we require that $Z_{top}$ and $Z_{bot}$ are
consistent within fluctuations for each accepted event. In practice,
this means that particles with $Z_{bot} < Z_{top}$ are rejected. In
almost all cases these are particles that have undergone a nuclear
spallation while traversing the detector material between top and
bottom of the instrument (for numerical detail, see section 5 and
figure \ref{interactions}).

The procedure to select a sample of a given element for further
analysis is as follows: first, with a very tight cut on $Z_{bot}$,
clean, penetrating nuclei are selected, and their distribution in
$Z_{top}$ is investigated. This yields the values for charge
resolution mentioned above, and it indicates the efficiency if a
charge cut on $Z_{top}$ is made. If the cut on $Z_{top}$ is of the
order of $\Delta{Z}_{top} \ge \pm 0.5$ charge units, the efficiency is
quite high. For certain elements, a tighter cut on $Z_{top}$ must be
made, and the efficiency is then reduced. Numerical details are given
in section 5 and table \ref{tab:eff}. It should be noted that the
tight cut on $Z_{bot}$ is only applied in order to determine the form
of the charge distribution on $Z_{top}$. For the final analysis, only
a very loose cut on $Z_{bot}$ is used to remove interacting
particles. This cut leads to a negligible reduction in overall charge
selection efficiency. 

Finally, it should be noted that ``edge'' effects can compromise the
charge resolution. To avoid this problem, tracks with an impact point
within 1 cm of the wavelength-shifter bars are excluded. This
creates a dead area in the detector of about 4 \% of the total area.

\subsection{Energy Analysis}
\label{sec:energy}

\subsubsection{Overview}

The energy of each cosmic-ray nucleus is obtained from the combined
signals of the Cherenkov counter and of the proportional tubes. The
top four double layers of tubes measure the specific ionization of
each particle, $\langle dE/dx \rangle$, while the bottom four double
layers detect transition radiation, measuring $\langle dE/dx + TR
\rangle$. While the main objective of the Cherenkov counter is to
identify low energy particles, below the level of minimum ionization,
the rapid increase of the signal with energy provides a good energy
measurement up to about 3 GeV per amu.

At energies between minimum ionization and the onset of TR (3 to 400
GeV per amu) the signals in the dE/dx tubes and the TR tubes are the
same and increase logarithmically with energy. This increase is slow
and can be used for an energy measurement unless it is obscured by
statistical signal fluctuations. The relative level of the
fluctuations decreases with the nuclear charge as $1/Z$. Therefore,
the energy measurement improves with increasing $Z$. Note that each
individual proportional tube provides an estimate for the specific
ionization by measuring the ratio
$\Delta{E}_{i}/\Delta{x}_{i}$. Hence, for any particle up to 16 tubes
provide independent measurements, which are required to be consistent
within fluctuations. For the final analysis, the average $\langle
dE/dx \rangle$, as defined in equation \ref{dedxeq} and
\ref{dedxtreq}, is used.

At the highest energies ($>$ 400 GeV per amu), the signals from the
dE/dx tubes and from the TR tubes diverge, and the rapid increase of
the TR signals with particle energy provides an excellent energy
measurement. However, particles in this energy region are extremely
rare, less abundant than particles in the minimum-ionization region by
more than four orders of magnitude. To uniquely identify these
particles, it is necessary but not sufficient to just require that
low-energy nuclei are rejected on the basis of their Cherenkov
signals. In addition, we require that the measurement of their
ionization energy loss $\langle dE/dx \rangle$ (performed with four
double layers of dE/dx tubes) places them at an energy level well
above minimum ionization (for details, see section~\ref{sec:hep} and
table~\ref{tab:flux1}). Thus, the combination of dE/dx tubes with the
TRD tubes is crucial for the success of the TRACER measurement at the
highest energies.

\subsubsection{Identification of Sub-Relativistic Particles ($<$ 3 GeV/amu)}
\label{sec:lep}

\begin{figure}[h]
\begin{center}
\includegraphics[width=0.99\textwidth]{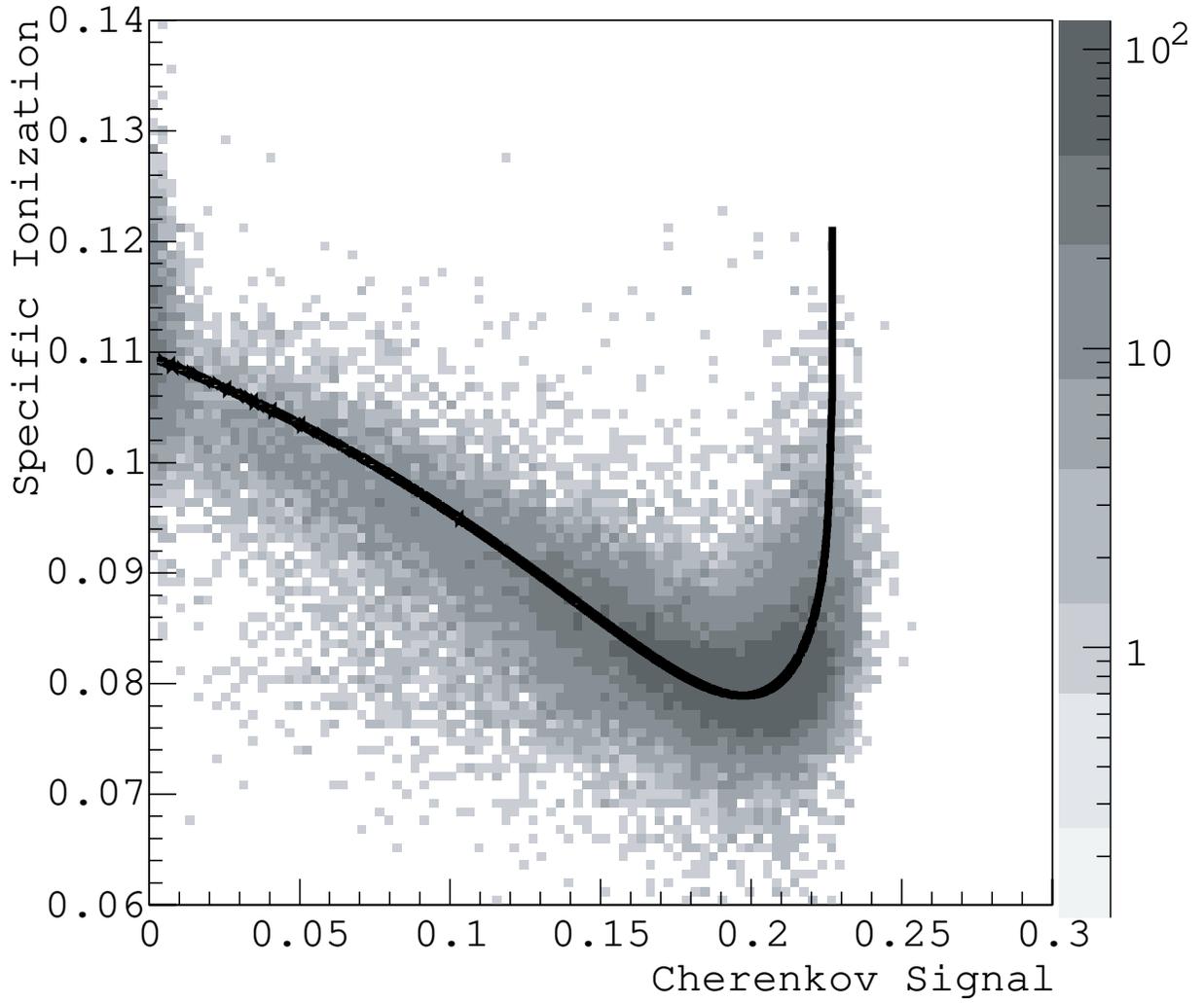}
\end{center}
\caption{Scatter plot of dE/dx vs. Cherenkov signals for iron
nuclei. The black line is the average response obtained from
simulations. The gray scale is logarithmic.}
\label{CerdEdx}
\end{figure}

To identify low energy particles, the signals from the Cherenkov
counter are compared with those of the dE/dx tubes. The dE/dx
response, shown in figure \ref{resp}, is well described by the
Bethe-Bloch formula. The Cherenkov counter response (figure
\ref{resp}) must be slightly modified to account for effects of delta
rays generated by the primary particle in the instrument, while the
signals from the proportional tube array remain unaffected by delta
rays. These effects are well understood and have been previously
studied and reported \citep{gahbauer03b,romero05b}. The scatter plot
in figure \ref{CerdEdx} shows the correlation of the ionization and
Cherenkov signals measured for iron nuclei. The solid line represents
the average response expected from simulations. The Cherenkov signal
increases with energy until it reaches saturation, while the
ionization signal clearly exhibits the level of minimum ionization
($\sim$ 3 GeV per amu). Note that the gray scale of the scatter plot
is logarithmic. Plots like this one for each charge exhibit the level
of fluctuations in the Cherenkov signal: it is about 8 \% for O nuclei
and 2.5 \% for Fe nuclei. A cut in the Cherenkov signal is placed at a
level that corresponds to the minimum in the ionization signal. Events
with Cherenkov signals below this cut are used to generate their
energy spectrum from 0.8 to 2.3 GeV per amu. Events with Cherenkov
signals above the cut are used to construct the energy spectra above
12 GeV per amu as described in the following section.

\subsubsection{High Energy Particles}

\label{sec:hep}

\begin{figure}[h]
\begin{center}
\includegraphics[width=.99\textwidth]{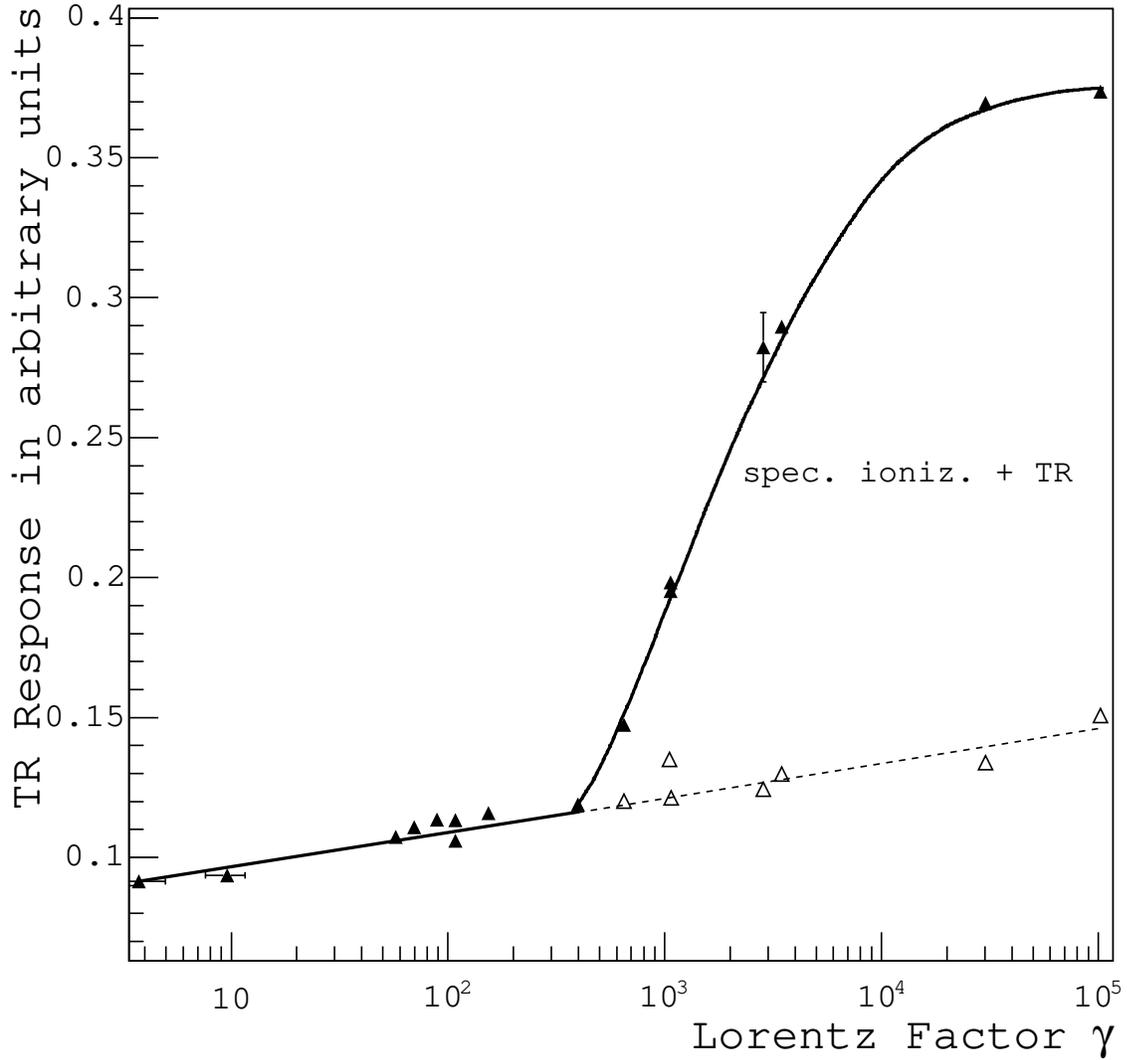}
\end{center}
\caption{Energy response of the Transition Radiation Detector as
measured at accelerators. Except for two highest-energy data points,
these measurements have been published previously
\citep{lheureux90}. The open symbols and dashed lines refer to
measurements of dE/dx only.}
\label{TRcurve}
\end{figure}

After removing all particles below about 3 GeV per amu, the rare high
energy particles are identified from the combination of the signals
from the dE/dx tubes and the TRD. These detectors have been calibrated
at accelerators using singly charged particles. Figure \ref{TRcurve}
shows the measured response for the TRD of the CRN instrument
\citep{lheureux90}.  The radiator combination and gas mixtures of
TRACER are identical to those of the CRN instrument. However, CRN used
plane multiwire proportional chambers rather than the layers of
cylindrical proportional tubes of TRACER. We have ascertained in
Monte-Carlo simulations that the accelerator calibrations for CRN remain
valid for the detector geometry of TRACER. Most of the data shown in
figure \ref{TRcurve} have been published previously
\citep{lheureux90}.

\begin{figure}[h]
\begin{center}
\includegraphics[width=.99\textwidth]{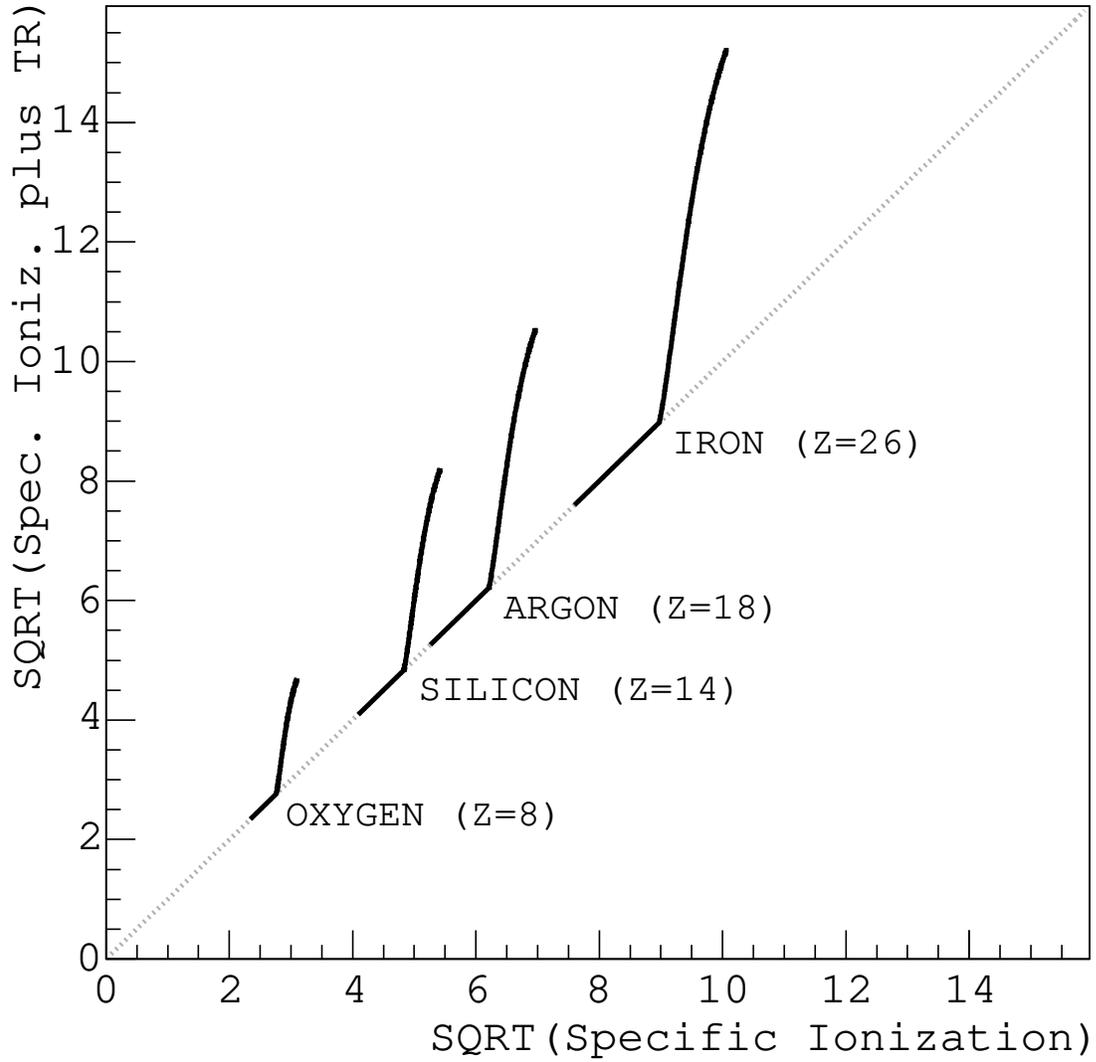}
\end{center}
\caption{Correlation of average responses (square root) of Transition
Radiation and specific ionization detectors for relativistic
nuclei. Four elements are displayed to illustrate the charge
dependence of the responses.}
\label{TRDvsDEDX}
\end{figure}

Below 400 GeV/amu we expect no observable transition radiation. Both
signals, $\langle dE/dx \rangle$ and $\langle dE/dx + TR \rangle$,
increase logarithmically with energy and will cluster along the
diagonal in a correlation plot of TRD response vs ionization response
as illustrated in figure \ref{TRDvsDEDX}. Above 400 GeV per amu, TR becomes
observable, and the signal from the TRD will increase above the
ionization signal. This manifests itself as a deviation from the
diagonal in the correlation plot. 

\begin{figure}[h]
\begin{center}
\includegraphics[width=0.90\textwidth]{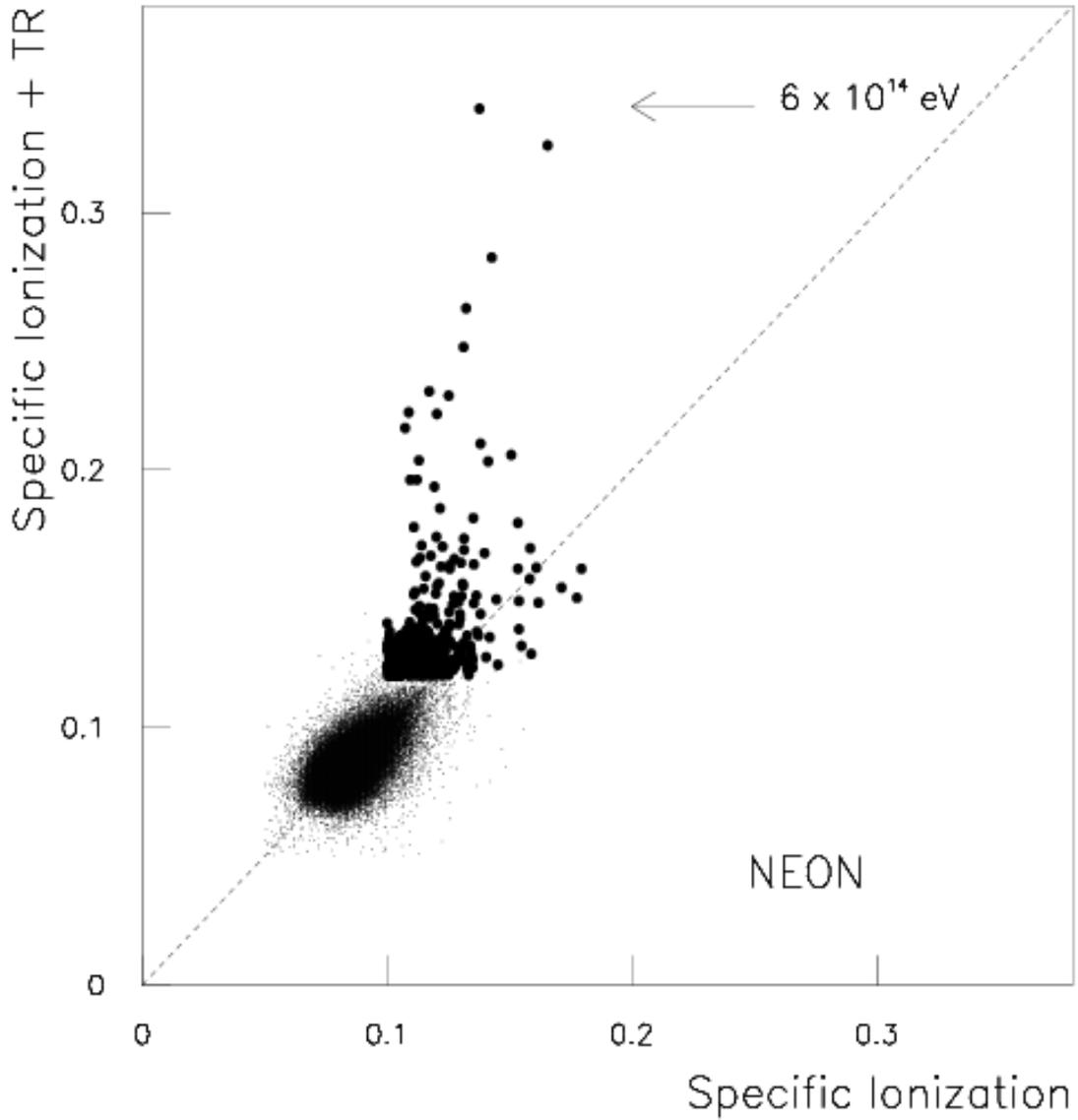}
\end{center}
\caption{Scatter plot of TR vs. dE/dx signal for neon nuclei. The
highlighted points represent the highest energy events measured with
the TRD. As expected the transition radiation events have signals in
the dE/dx detector which are well above the minimum ionization
level. Units are arbitrary.}
\label{TRdEdx}
\end{figure}

As an example, figure \ref{TRdEdx} shows the observed cross
correlation between the TRD and the ionization signals for neon nuclei
($Z$ = 10) above minimum ionization. The small black points represent
the numerous events with energies below the onset of TR. The rare high
energy particles with clear TR signals are highlighted. As expected,
the data follow the response illustrated in figure
\ref{TRDvsDEDX}. Note that the highest energy events (the $TR$
events) stand out without any background in other regions of the
scatter plot. The most energetic neon nucleus in this sample of data
has an energy of $6 \times 10^{14}$eV.

We must emphasize again that the off-diagonal position of these higher
energy events defines them uniquely as TR-events for the selected
charge (Z=10 in figure \ref{TRdEdx}). Any spillover due to
misidentified charges, with Z either larger or smaller than the
selected value, would lead to signals along the diagonal, but would
not contaminate the well defined ``TR-tails''.

\section{Absolute Flux Measurements}
\label{sec:flux}
 

Each event that passes the data analysis cuts is classified in energy
as either as a \emph{Cherenkov Event} ($<$ 3 GeV per amu),
\emph{dE/dx Event} (10 - 400 GeV per amu) or \emph{TR Event} ($>$ 400
GeV per amu). Events are sorted into energy bins of width
$\Delta{E}_i$ and a differential energy spectrum is constructed for
each elemental species. We present the spectra in terms of an absolute
flux $dN/dE$ at the top of the atmosphere. To convert from the number of
events $\Delta{N}_i$ in a particular energy bin $\Delta{E}_i$ to an
absolute differential flux $\mathrm{d}N/\mathrm{d}E (i)$ one must
compute the exposure factor, effective aperture, efficiency of the
cuts, and unfold the instrument response:

\begin{equation}\label{dNdE}
\frac{\mathrm{d}N}{\mathrm{d}E}(i)= \frac{\Delta N_i}{\Delta E_i}  \cdot \frac{1}{T_l} \cdot  \frac{1}{\varepsilon_i} \cdot \frac{1}{A_i} \cdot C_i
\end{equation}

with $\Delta{N}_i$ the number of events, $\Delta{E}_i$ the energy
range, $T_l$ the live-time, $\varepsilon_i$ the efficiency of analysis
cuts, $A_i$ the effective aperture and $C_i$
the ``overlap correction'' due to misidentified events from
neighbouring energy bins. For Cherenkov events one also must take into
account that $\Delta{E}_i$, the energy interval on top of the
atmosphere, is not equal to the measured energy interval because of
energy losses in the atmosphere and in the detector.

\emph{Livetime} --- The live time is determined by comparing the raw
trigger rate with the number of events stored on disk. With a raw
trigger rate of 60 Hz, the dead time was negligible throughout the
flight, and the total live time is 787,200 seconds.

\emph{Tracking Efficiency} --- The efficiency of the trackfit is 95 \%
 for all charges, and is independent of energy. Included is a consistency check in the signal of the
 tubes contributing to the $\langle dE/dx \rangle$ and $\langle dE/dx
 + TR \rangle$ measurements. 

\emph{Charge Efficiency} --- For the more abundant elements (O, Ne,
Mg, Si) a charge cut of $Z_{top} \pm$ 0.5 is made. The efficiencies
range from 85 \% to 72 \% (O, Si) and are listed in table
\ref{tab:eff}. For the less abundant elements (S, Ar, Ca) a more
conservative cut of $Z_{top} \pm$ 0.1 is imposed to avoid
contamination from adjacent charges, which yields a reduced efficiency of
17 \%. For iron, a cut of $Z_{top} \pm$ 1.0 is used at an efficiency
of 90 \%. The consistency cut on the charge at the bottom of the
instrument $Z_{bot}$ can be sufficiently loose to not affect the
overall efficiency.

\begin{figure}[h]
\begin{center}
\includegraphics[width=0.9\textwidth]{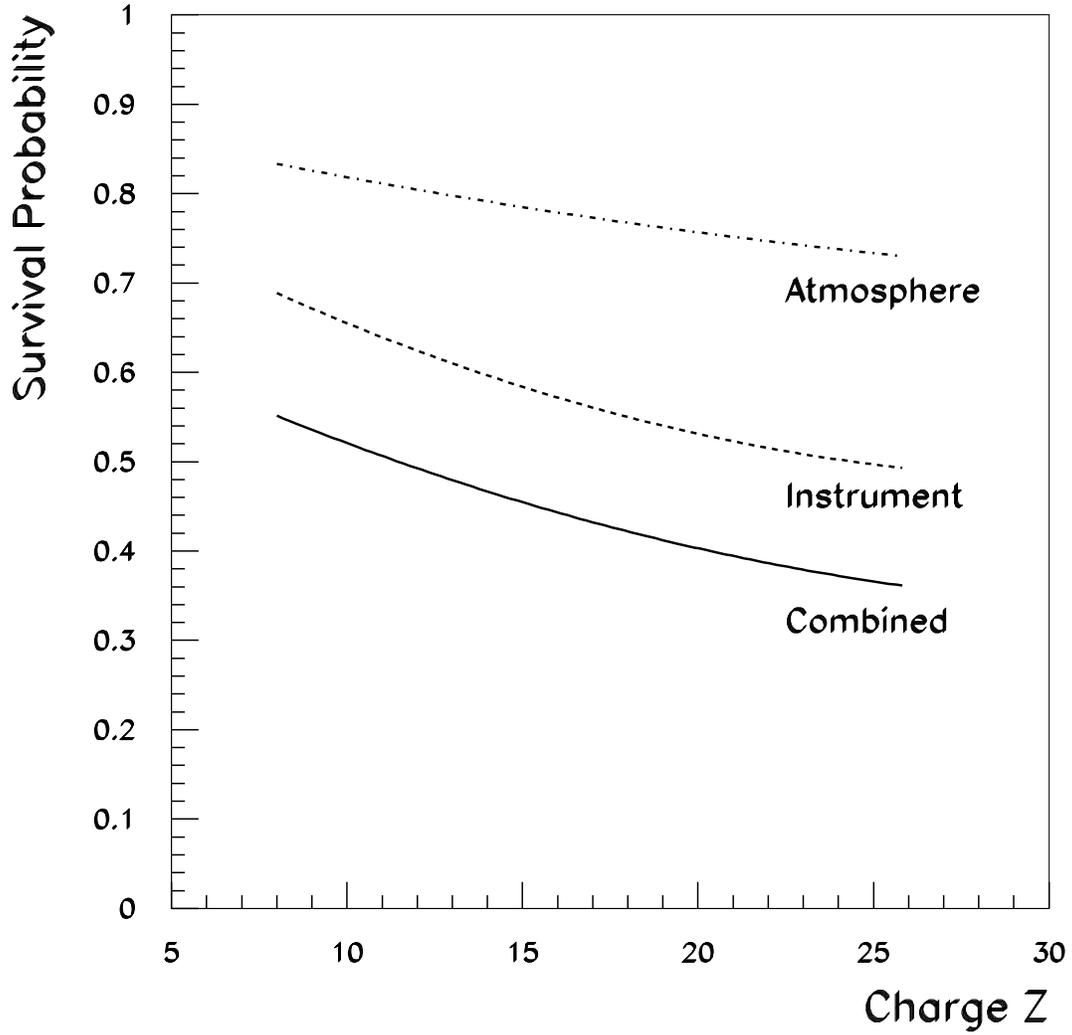}
\end{center}
\caption{Survival probability all nuclei incident at a zenith angle of
30$^\circ$ in a residual atmosphere of 3.91 g/cm$^2$ (dash-dot),
within the TRACER instrument (dash), and total survival probability
(solid).}
\label{interactions}
\end{figure}

\emph{Effective Aperture} --- The overall geometric factor of TRACER
is 5.04 m$^2$ sr. For the flux computation an effective aperture $A_i$
is calculated, which includes the geometric factor, losses due to
nuclear spallations, and ``dead'' regions in the detector.

\begin{equation}\label{geometry}
A_i = A \cdot 2 \pi \int^{\pi /2}_{\theta=0}P_I(\theta)P_D(\theta)\cos \theta \ \mathrm{d}(\cos \theta)
\end{equation}

with $A$ the area of the detector (206 cm $\times$ 206 cm),
$P_I(\theta)$ the probability of survival against spallation in the
atmosphere and in the instrument as a function of zenith angle
$\theta$ (see figure \ref{interactions}), and $P_D(\theta)$ the
probability that a particle passing through the top scintillator will
traverse the entire instrument without encountering any insensitive
detector regions. The quantity P$_{D}(\theta)$ is determined from a
Monte Carlo simulation. Insensitive detector regions lead to a
reduction in the overall geometric factor by about 10~\%.

Losses due to nuclear interactions are computed using a combination of
measured charge-changing cross sections and parameterizations obtained
by \citet{heckmann78} and \citet{westfall79}. The interaction losses
are taken to be independent of energy but do depend on the particle
mass. Interactions in the atmosphere are determined by measuring the
residual atmosphere above the payload, which for this flight was
monitored by pressure sensors and GPS modules attached to the
instrument. The average residual atmosphere was 3.91
g/cm$^2$. Interaction probabilities in the instrument are obtained by
numerical computation using a complete list of materials in
TRACER. The fraction of interacting nuclei in the instrument can also
be obtained experimentally from a comparison of the reconstructed
nuclear charges for each element in the top and bottom
scintillators. Both methods lead to consistent results. As an example,
an oxygen or iron nucleus at an incident angle of 30$^\circ$ has a
probability of survival in the atmosphere of 83~\% or 73\% and
probability of survival in the instrument of 70~\% or 49~\%,
respectively (see figure \ref{interactions} for all elements).

\begin{figure}[h]
\begin{center}
\includegraphics[width=0.9\textwidth]{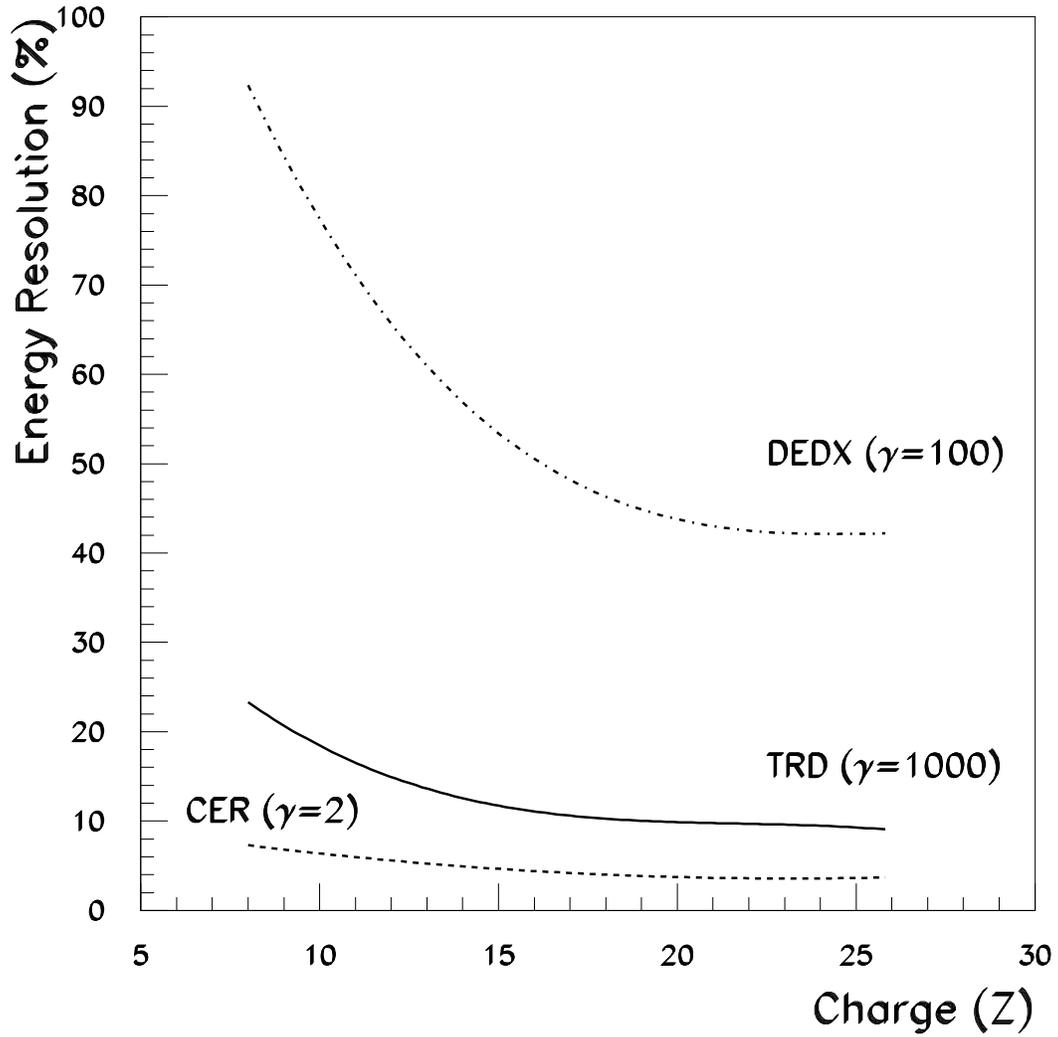}
\end{center}
\caption{Energy resolution (1$\sigma$) of individual detector
subsystems vs charge $Z$, and for typical energies: Cherenkov:
$\gamma$ = 2 (dashed line), dE/dx: $\gamma$ = 100 (dash-dot) and TRD:
$\gamma$ = 1000 (solid line)}
\label{eres}
\end{figure}

\emph{Energy Overlap Corrections} --- As TRACER has three detectors, each
with a different energy response and energy resolution (see
figure~\ref{resp} and figure~\ref{eres}), the widths of the energy
intervals into which the data are sorted must be appropriately
chosen. The intrinsic relative signal fluctuations decrease with
increasing $Z$, essentially as $1/Z$ (figure~\ref{eres}). Therefore,
the number of acceptable energy bins increases for elements with
higher charge number.

As the shape of the energy spectra is not expected to vary rapidly
with energy, choosing too fine a subdivision of the energy scale might
just lead to increased statistical uncertainty in each energy bin, and
hence, not improve the quality of the results. More importantly, the
widths of the bins must be commensurate with the energy resolution of
the detector. Otherwise, overlap corrections for events being
misinterpreted from a neighbouring energy bin would become substantial
and could possibly bias the result for a steeply falling spectrum.

In the current analysis of the TRACER data, the bin widths are chosen
fairly wide, such that the overlap correction factor can be determined
with the help of an iterative Monte-Carlo simulation. The corrections
are computed for each energy bin and element. They are fairly small,
in general $\le 20~\%$ (i.e. $0.8 \leq C_i \leq 1.2$). 
 
\section{Measured Energy Spectra}
\label{sec:spectra}



The analysis yields, for each species, numbers of events $\Delta{N}_i$
sorted into energy bins $\Delta{E}_i = E_{i+1} - E_{i}$. The ratio
$\Delta{N}_{i}/\Delta{E}_{i}$ defines the differential intensity $dN/dE$
for this interval:

\begin{equation}\label{dndei}
\frac{dN}{dE}(i) = \frac{\Delta{N}_{i}}{\Delta{E}_{i}}
\end{equation}

We assume that the energy spectrum is represented by a power law
$CE^{-\alpha}$ over this interval, such that:

\begin{equation}\label{dn}
\Delta{N}_{i} = \int^{E_{i+1}}_{E_{i}} C E^{-\alpha} dE
\end{equation}

We then plot the spectral value $\frac{dN}{dE}(i)$ at an energy level
$\hat{E}$ defined such that:

\begin{equation}\label{dnde}
\frac{dN}{dE} = C \cdot \hat{E}^{-\alpha},
\end{equation}

\begin{equation}\label{ehat}
\hat{E} = \frac{1}{(E_{i+1} - E_{i})} \quad \frac{1}{(1-\alpha)} \quad (E_{i+1}^{1-\alpha} - E_{i}^{1-\alpha})^{-1/\alpha}
\end{equation}

Since $\hat{E}$ depends on the value of $\alpha$, an iterative
procedure has been used. It is found that the dependence of $\hat{E}$
on $\alpha$ is not very strong.

For the spectra presented in this work, the highest energy intervals
are integral bins, with $E_{i+1} \rightarrow \infty$. We then plot the
data point at the median energy for this bin, $\hat{E}_{int}$:


\begin{equation}\label{eint}
\hat{E}_{int} = [E^{1-\alpha}_{i}/2]^{1/1-\alpha} 
\end{equation}

The value of the flux at this energy is:


\begin{equation}\label{fint}
\left| \frac{dN}{dE} \right| _{int} = \frac{\Delta{N}_{i}(\alpha-1)}{ E^{1-\alpha}_{i}} \cdot \hat{E}_{int}^{-\alpha}
\end{equation}

\begin{figure}[t]
\begin{center}
\includegraphics[width=.90\textwidth]{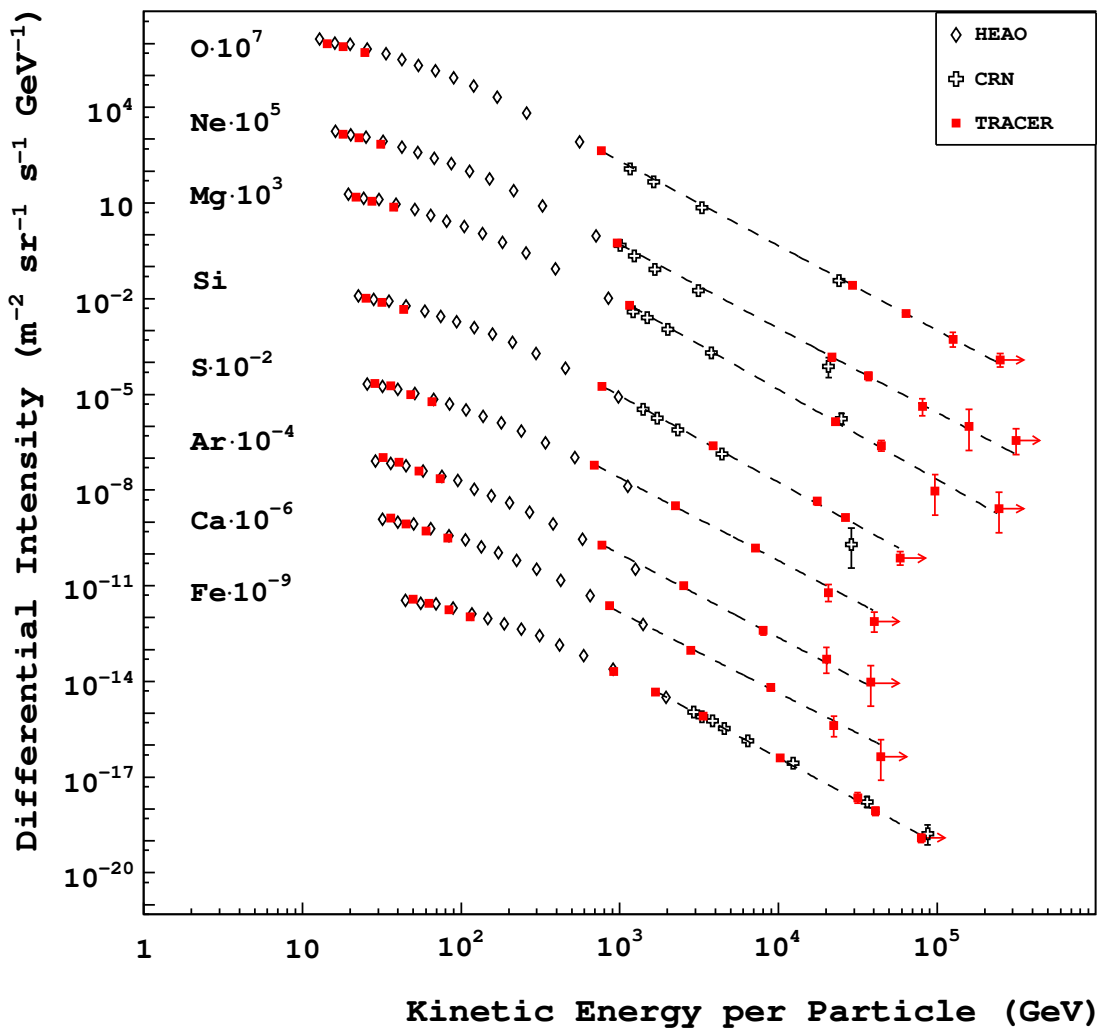}
\end{center}
\caption{Differential energy spectra vs energy per particle of the
cosmic ray nuclei : O, Ne, Mg, Si, S, Ar, Ca and Fe. Results from the
TRACER 2003 flight are indicated by the solid squares. Existing data
from the HEAO-3 experiment (open diamonds : \citet{engelmann90}) and
the CRN experiment (open crosses : \citet{muller91}) are shown for
comparison. The dashed line represents an independent power-law fit to
each spectrum above 20 GeV per amu.}
\label{tracerspec}
\end{figure}

The energy spectra, in terms of absolute intensities, for the elements
O, Ne, Mg, Si, S, Ar, Ca, and Fe are presented in tables
\ref{tab:flux1} and \ref{tab:flux2} and are plotted in figure
\ref{tracerspec}. The low energy data points ($\le 2.3$ GeV per amu)
come from an analysis of the Cherenkov counter, using a restricted
angular acceptance. Note that the energy in the table is given in
units of kinetic energy per amu, whereas in figure \ref{tracerspec}
the units are given in terms of kinetic energy per particle. Data from
the TRACER 2003 flight are indicated by the solid squares. For clarity
the intensity of each element is scaled by a factor shown on the
left. Existing data from measurements in space with HEAO-3
\citep{engelmann90} and CRN \citep{muller91} are shown for
comparison. In the energy regions where overlap exists, the agreement
of the results appears to be quite good.

\begin{figure}[ht]
\begin{center}
\includegraphics [width=0.90\textwidth]{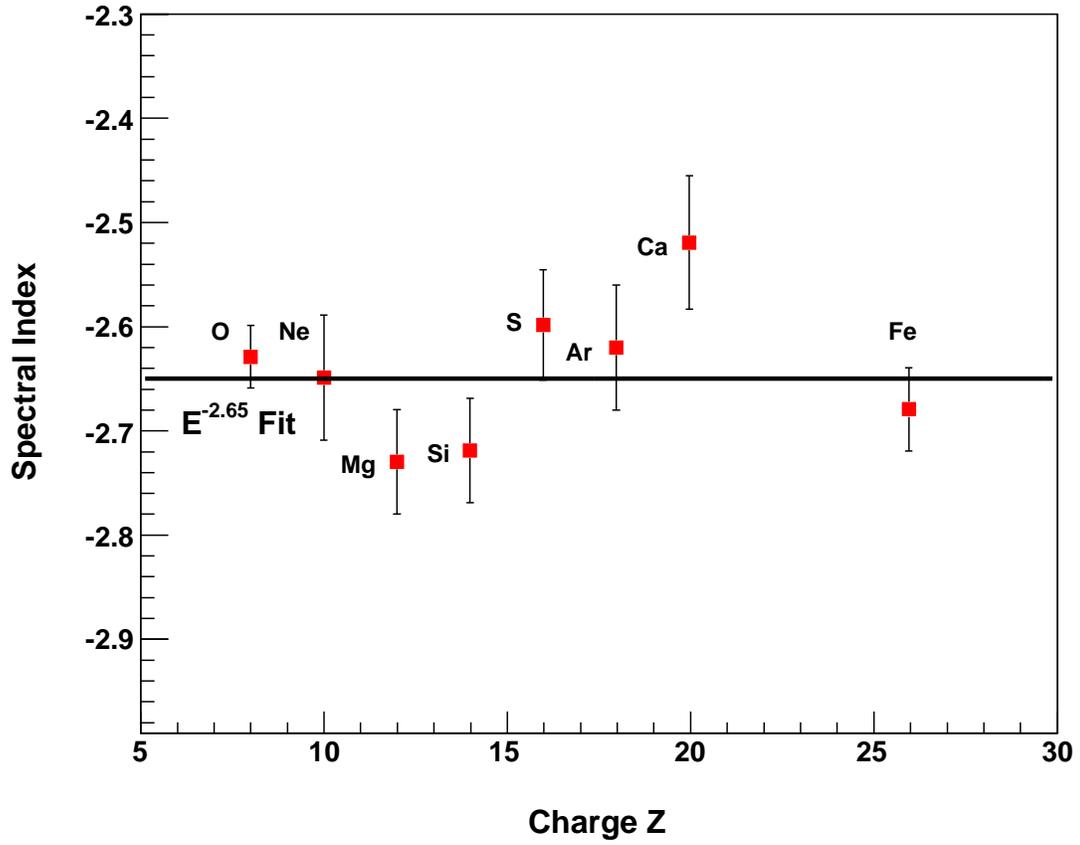}
\end{center}
\caption{Spectral indices of a best power-law fit to the combined
TRACER and CRN data above 20 GeV per amu. The line indicates the an
average spectral fit of E$^{-2.65}$.}
\label{powerlaw} 
\end{figure}

We note the large range in intensity (ten decades) and particle energy
(four decades) covered by TRACER.  As can be seen, the energy spectra
extend up to and beyond 10$^{14}$ eV. The eight elements for which
spectra are shown in figure \ref{tracerspec} represent the major
primary species from oxygen to iron; contributions to these species
from spallation produced secondary elements are expected to be
small. We may fit the high-energy part ($>$ 20 GeV per amu) of each
spectrum to a power law in energy $E^{-\alpha}$. The resulting
spectral indices are shown in figure \ref{powerlaw}. Remarkably, there
is no significant trend of the spectral indices with charge $Z$, and
all indices fit well to an average of $\alpha$ = 2.65 $\pm$ 0.05. It
must be emphasized that this result is not an artifact of the method
of defining the differential intensities described in the previous
sections. The similarity of the spectral shapes must indicate that
acceleration and propagation mechanisms are essentially the same for
all species. However, the statistical errors do not exclude the
possibility of deviations from a pure power-law behavior for
individual spectra.


\section{Comparison with other data}
\label{sec:otherdata}

\begin{figure}[h]
\begin{center}
\includegraphics [width=0.90\textwidth]{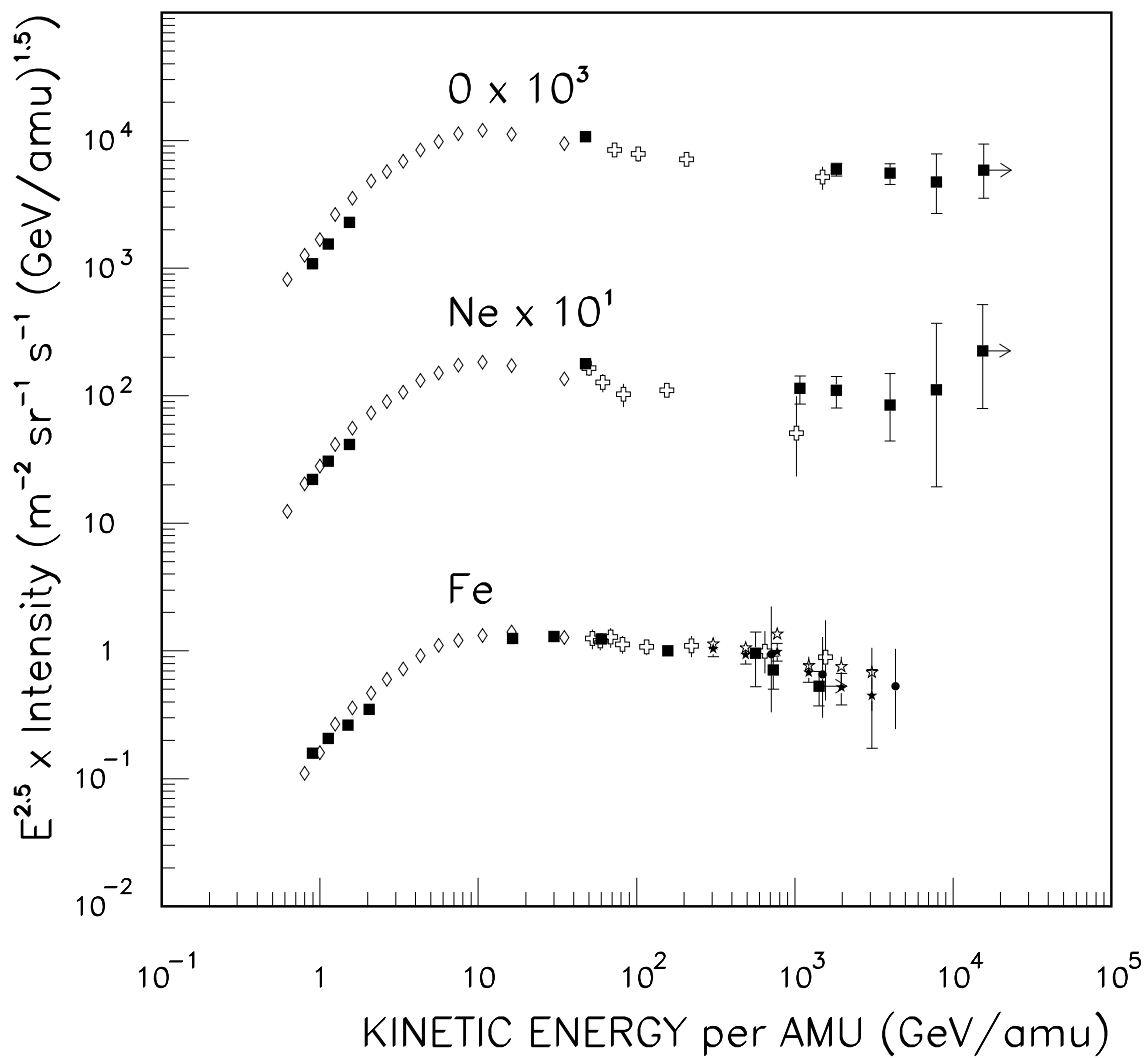}
\end{center}
\caption{Energy spectra multiplied with $E^{2.5}$ for O, Ne, and Fe,
for different observations: TRACER (solid squares), HEAO-3 (diamonds:
\citet{engelmann90}), CRN (crosses: \citet{muller91}), RUNJOB (filled
circles: \citet{derbina05}) and HESS (open stars: QGSJET model, filled
stars: SYBILL model; \citet{aharonian07a}). The error bars shown are
statistical.}
\label{fe-ne-o} 
\end{figure}

In figure \ref{tracerspec}, the energy spectra obtained with TRACER
are shown together with measurements in space with the HEAO-3
satellite up to $ \sim $ 40 GeV per amu, and with the CRN detector on
the Space Shuttle up to about 1000 GeV per amu. However, a
multi-decade plot such as that of figure \ref{tracerspec} may obscure
the finer details in a steeply falling spectrum. To illustrate this,
we show the energy spectra for O, Ne, and Fe, in figure \ref{fe-ne-o}
multiplied with $E^{2.5}$. It is apparent that the differences between
individual data sets seem to be mostly statistical.

We note that figure \ref{fe-ne-o} also includes a recent determination
of the high-energy iron spectrum that was performed with the HESS
Imaging Air Cherenkov Telescope on the ground. Here it was possible
for the first time to observe directly the Cherenkov light of
iron-group nuclei in the atmosphere before they interact
\citep{aharonian07a}. Again, the agreement with the TRACER data is
good, although uncertainty about the choice of a nuclear interaction
model affects the interpretation of the HESS measurement. 

There are also several recent or current balloon-borne detectors that
are designed to obtain measurements of the high-energy cosmic-ray
composition, for instance RUNJOB \citep{derbina05}, ATIC
\citep{panov06}, and CREAM \citep{seo06}; all of these employ mostly
calorimetric techniques. The RUNJOB experiment has published energy
spectra for mostly elemental groups. The iron spectrum from this
experiment is also included in figure \ref{fe-ne-o}. Again we see that
there is good agreement with the TRACER data. Spectra for a few
individual heavy elements, more limited in energy coverage and
statistics than TRACER, have been reported for ATIC and CREAM at
conferences \citep{panov06,ahn07, zei07}.  These results are not
included here as they are still labeled preliminary. However, the
agreement in the regions of overlap appears to be quite good.

\begin{figure}[h]
\begin{center}
\includegraphics [width=0.90\textwidth]{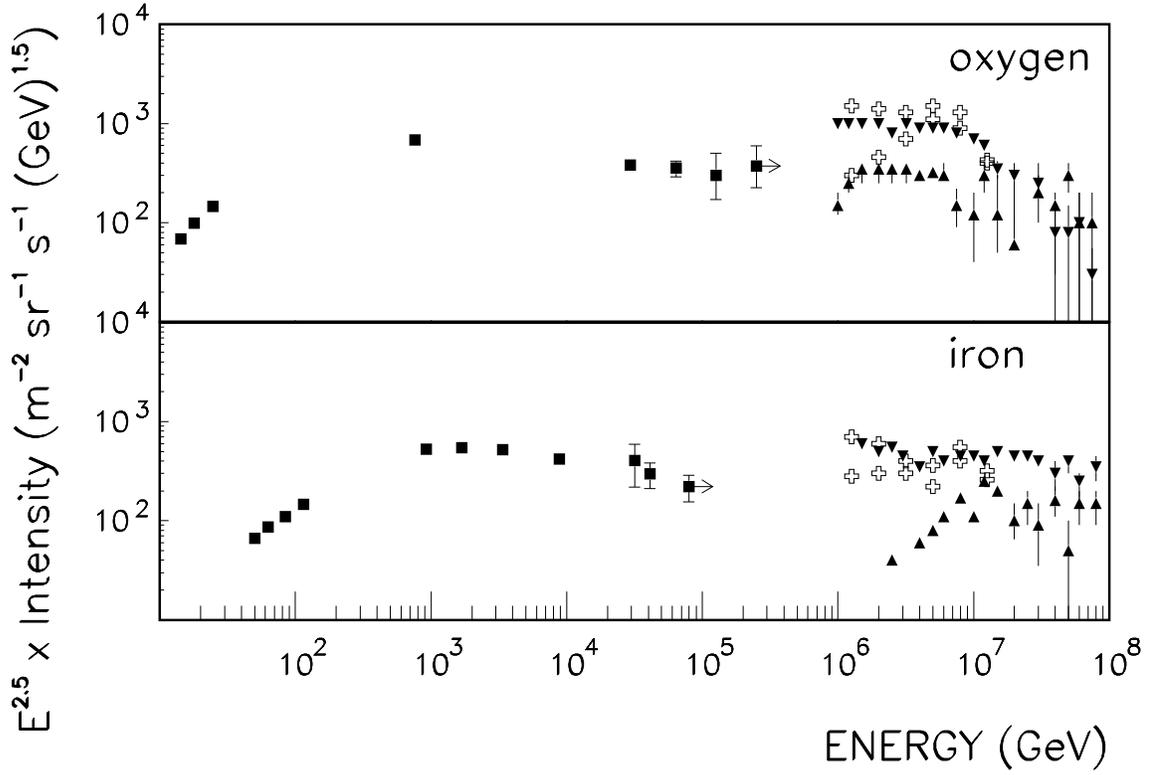}
\end{center}
\caption{Energy Spectra from TRACER (solid squares) compared with the
interpretation of air shower data of KASCADE (solid triangles, for two
different interaction models : \citet{antoni05}) and of EAS-TOP (open
crosses, two data points for each energy represent upper and lower
limits: \citet{navarra03}). The spectra are for oxygen and for iron
for TRACER, but for the ``CNO-group'' and the ``Fe-group'' for the
other observations.}
\label{airshower} 
\end{figure}

Data for oxygen and iron from TRACER are compared in figure
\ref{airshower} with spectra derived from indirect air-shower
observations by the EAS-TOP collaboration \citep{navarra03}, and by
the KASCADE group for two different nucleus-nucleus interaction models
\citep{antoni05}. These groups do not report results for individual
elements: the fluxes for the ``CNO group'' most likely have about twice
the intensity than oxygen alone, while the ``iron group'' probably is
dominated by iron. Our results do not yet overlap with the energy
region of the air shower data, but the gap is becoming smaller, in
particular for oxygen. Additional measurements should indeed lead to
significant constraints on the air shower interpretations.

\section{Conclusions}
\label{sec:conclusions}

The energy spectra of the cosmic-ray nuclei determined in this
investigation extend to energies around $10^{14}$ eV per particle, or
significantly higher for the more abundant nuclei. We believe that
this data set represents the most comprehensive information about the
spectra of the heavier primary nuclei currently available. 

We have studied in some detail how these results provide constraints
on our current understanding of the origin of galactic cosmic rays and
of models of acceleration and propagation. We realize that the upper
energy limit of the measurement is still below the region of the
cosmic-ray ``knee'', so perhaps it is not surprising that no
significant changes in cosmic-ray composition are yet
observed. However, the great similarity between the individual energy
spectra is remarkable. The result of our studies of these features is
currently being prepared for a separate publication.

It is clear that an extension of the measurements to still higher
energy is desirable. The dynamic range of the TRD used here is not
exhausted with the current measurement. Hence, in order to reach
higher energy, the TRACER detector would either have to be increased
in size, which seems impractical as the instrument is already the
largest balloon-borne cosmic-ray detector in existence, or would have
to be subjected to additional long-duration balloon flights. Ideally,
an instrument like TRACER should be exposed in space for several
years.

An important objective for measurements beyond the current TRACER
results would be an extension of the dynamic range of the detector,
such that the light secondary elements (Li, Be, B) as well as carbon
and nitrogen can be covered together with the heavier nuclei. If
successful to sufficiently high energies ($\sim 10^3$ GeV per amu),
this measurement would lead to a determination of the L/M abundance
ratio, i.e., to a determination of the propagation pathlength at high
energy. 

This would be an essential step towards a better understanding of the
galactic propagation of cosmic rays. To make this measurement
possible, the electronics of TRACER has recently been completely
refurbished, and another successful long-duration balloon flight was
performed in 2006. The analysis of these new data is currently in
progress and results will be reported in due course.

\acknowledgements

We gratefully acknowledge the contributions of Drs. German Hermann
and Scott Wakely to the design and construction of TRACER. We
appreciate the services of the University of Chicago Engineering
Center, especially Gary Kelderhouse, David Pernic and Gene Drag. We
thank the staff of the Columbia Scientific Balloon Facility, especially
David Sullivan, and the National Science Foundation and United States
Antarctic Program for support in conducting this flight. We are
indebted to Scott Cannon for his contribution in the development of
the thermal protection for the instrument. This work was supported by
NASA grants NAG5-5305, NN04WC08G and NNG06WC05G. MI acknowledges the
Grant-in-Aid for Scientific Research of the Japan Society for the
Promotion of Science (JSPS), No. 17540226. Numerous students have
participated in the construction of the instrument under support from
the Illinois Space Grant Consortium.



\clearpage

\begin{table}
\caption{\label{tab:eff} Charge selection efficiencies}
\begin{center}
\begin{tabular}{lcccccccc} 
Element  & {\bf 8} & {\bf 10}& {\bf 12} & {\bf 14} & {\bf 16} & {\bf 18}& {\bf 20} & {\bf 26}\\ 
  Cut (Charge Units) & $\pm$0.5 & $\pm$0.5 & $\pm$0.5 & $\pm$0.5 & $\pm$0.1 & $\pm$0.1 & $\pm$0.1 & $\pm$1.0 \\
  Efficiency & 0.89 & 0.83 & 0.75 & 0.72 & 0.17 & 0.18 & 0.18 & 0.90 \\
\end{tabular}
\end{center}
\end{table}

\clearpage


\newcommand{\ibin}{b}

\begin{table}
\begin{minipage}[h]{20cm}

\caption{\label{tab:flux1} {\bf Measured Intensities (Oxygen - Silicon)}}

\footnotesize

\begin{tabular}{lr@{ - }lrrr@{}l@{$\times$}l} \hline 
{\bf Element}  & \multicolumn{2}{c} {\bf Energy Range} & \multicolumn{1}{c}{\bf Kinetic Energy\tablenotemark{1}} & \multicolumn{1}{c}{\bf Number of} &  \multicolumn{3}{c} {\bf  $\quad$ Differential Intensity}\\
 & \multicolumn{2}{c} {\bf (GeV/amu) } & \multicolumn{1}{c}{\bf \^{E} (GeV/amu)} & \multicolumn{1}{c}{\bf Events } & \multicolumn{3}{c} {\bf $\quad$ (m$^2$ s  sr GeV/amu)$^{-1}$}\\ \hline

\vspace{0.05cm} \\ 
 \bf{O   (Z = 8)} &       0.8 &        1.0 &        0.9$\qquad \quad$ &     25437 $\quad$ & $\qquad$(1.71 $\;$&$\pm$ 0.01)&$\;10^0$   \\
 &        1.0 &        1.3 &        1.1$\qquad \quad$ &     24237 $\quad$ & $\qquad$(1.38 $\;$ &$\pm$ 0.01)&$\;10^0$   \\
 &        1.3 &        1.9 &        1.5$\qquad \quad$ &     38226 $\quad$ & $\qquad$(8.95 $\;$ &$\pm$ 0.05)&$\;10^{-1}$   \\
 &       12 &      292 &       48$\qquad \quad$ &    456918 $\quad$ & $\qquad$(7.27 $\;$ &$\pm$ 0.01)&$\;10^{-4}$   \\
 &     1315 &     2627 &     1834$\qquad \quad$ &        77 $\quad$ & $\qquad$(4.4 $\;$ &$\pm$ 0.5)&$\;10^{-8}$   \\
 &     2627 &     6371 &     4001$\qquad \quad$ &        29 $\quad$ & $\qquad$(5.8 $\;$ &$\pm$ 1.1)&$\;10^{-9}$   \\
 &     6371 &     9834 &     7872$\qquad \quad$ &         5 $\quad$ & $\qquad$(9.1 $\;$ & $^{+ 6.0} _{- 3.9}$)&$\;10^{-10}$ \\ 
 &     9834 &            &    15611$\qquad \quad$ &         6 $\quad$ & $\qquad$(2.0 $\;$ &$^{+ 1.2} _{- 0.8}$)&$\;10^{-10}$  \\ 
\vspace{0.05cm} \\
 \bf{Ne   (Z = 10)} & $\quad$        0.8 &        1.0 &        0.9 $\qquad \quad$ &      4031 $\quad$ & $\qquad$(3.08 $\;$ &$\pm$ 0.05)&$\;10^{-1}$   \\
 & $\quad$        1.0 &        1.3 &        1.1 $\qquad \quad$ &      3790 $\quad$ & $\qquad$(2.42 $\;$ &$\pm$ 0.04)&$\;10^{-1}$   \\
 & $\quad$        1.3 &        1.9 &        1.5 $\qquad \quad$ &      5639 $\quad$ & $\qquad$(1.48 $\;$ &$\pm$ 0.02)&$\;10^{-1}$   \\
 & $\quad$         12 &        292 &         48 $\qquad \quad$ &     57991 $\quad$ & $\qquad$(1.21 $\;$ &$\pm$ 0.01)&$\;10^{-4}$   \\
 & $\quad$        895 &       1316 &       1080 $\qquad \quad$ &        16 $\quad$ & $\qquad$(3.2 $\;$ &$\pm$ 0.8)&$\;10^{-8}$   \\
 & $\quad$       1316 &       2627 &       1834 $\qquad \quad$ &        13 $\quad$ & $\qquad$(8.2 $\;$ &$\pm$ 2.3)&$\;10^{-9}$   \\
 & $\quad$       2627 &       6371 &       4001 $\qquad \quad$ &         4 $\quad$ & $\qquad$(8.9 $\;$ & $^{+ 6.9} _{- 4.2}$)&$\;10^{-10}$ \\
 & $\quad$       6371 &       9834 &       7872 $\qquad \quad$ &         1 $\quad$ & $\qquad$(2.2 $\;$ & $^{+ 5.0} _{- 1.8}$)&$\;10^{-10}$ \\
 & $\quad$       9834 &  &      15611 $\qquad \quad$ &         2 $\quad$ & $\qquad$(7.8 $\;$ & $^{+ 10.1} _{- 5.0}$)&$\;10^{-11}$ \\
\vspace{0.05cm} \\ 
 \bf{Mg   (Z = 12)} &$\quad$        0.8 &        1.0 &        0.9 $\qquad \quad$ &      4603 $\quad$ & $\qquad$(4.02 $\;$ &$\pm$ 0.06)&$\;10^{-1}$   \\
 & $\quad$        1.0 &        1.3 &        1.1 $\qquad \quad$ &      4184 $\quad$ & $\qquad$(2.96 $\;$ &$\pm$ 0.05)&$\;10^{-1}$   \\
 & $\quad$        1.3 &        1.9 &        1.5 $\qquad \quad$ &      6339 $\quad$ & $\qquad$(1.89 $\;$ &$\pm$ 0.02)&$\;10^{-1}$   \\
 & $\quad$         12 &        292 &         48 $\qquad \quad$ &     59374 $\quad$ & $\qquad$(1.60 $\;$ &$\pm$ 0.01)&$\;10^{-4}$   \\
 & $\quad$        696 &       1316 &        946 $\qquad \quad$ &        23 $\quad$ & $\qquad$(3.6 $\;$ &$\pm$ 0.7)&$\;10^{-8}$   \\
 & $\quad$       1316 &       2627 &       1834 $\qquad \quad$ &         9 $\quad$ & $\qquad$(6.3 $\;$ & $^{+ 2.9} _{- 2.1}$)&$\;10^{-9}$ \\ 
 & $\quad$       2627 &       6371 &       4001 $\qquad \quad$ &         1 $\quad$ & $\qquad$(2.4 $\;$ & $^{+ 5.6} _{- 2.0}$)&$\;10^{-10}$ \\ 
 & $\quad$       6371 &  &      10113 $\qquad \quad$ &         1 $\quad$ & $\qquad$(6.7 $\;$ & $^{+ 15.4} _{- 5.5}$)&$\;10^{-11}$ \\ 
\vspace{0.05cm} \\ 
 \bf{Si   (Z = 14)} &$\quad$        0.8 &        1.0 &        0.9 $\qquad \quad$ &      3127 $\quad$ & $\qquad$(3.07 $\;$ &$\pm$ 0.06)&$\;10^{-1}$   \\
 & $\quad$        1.0 &        1.3 &        1.1 $\qquad \quad$ &      2812 $\quad$ & $\qquad$(2.27 $\;$ &$\pm$ 0.04)&$\;10^{-1}$   \\
 & $\quad$        1.3 &        1.9 &        1.5 $\qquad \quad$ &      4292 $\quad$ & $\qquad$(1.39 $\;$ &$\pm$ 0.02)&$\;10^{-1}$   \\
 & $\quad$         12 &         73 &         28 $\qquad \quad$ &     36369 $\quad$ & $\qquad$(5.23 $\;$ &$\pm$ 0.03)&$\;10^{-4}$   \\
 & $\quad$         73 &        292 &        138 $\qquad \quad$ &      2278 $\quad$ & $\qquad$(7.2 $\;$ &$\pm$ 0.2)&$\;10^{-6}$   \\
 & $\quad$        567 &        696 &        627 $\qquad \quad$ &        18 $\quad$ & $\qquad$(1.3 $\;$ &$\pm$ 0.3)&$\;10^{-7}$   \\
 & $\quad$        696 &       1316 &        946 $\qquad \quad$ &        24 $\quad$ & $\qquad$(4.0 $\;$ &$\pm$ 0.8)&$\;10^{-8}$   \\
 & $\quad$       1316 &  &       2088 $\qquad \quad$ &         6 $\quad$ & $\qquad$(2.2 $\;$ & $^{+ 1.3} _{- 0.9}$)&$\;10^{-9}$ \\

\end{tabular}
\tablenotetext{1}{see text for definition of $\hat{E}$}
\end{minipage}
\end{table}

\normalsize

\clearpage


\begin{table}
\begin{minipage}[h]{20cm}

\caption{\label{tab:flux2} {\bf Measured Intensities (Sulphur - Iron)}}

\footnotesize

\begin{tabular}{lr@{ - }lrrr@{}l@{$\times$}l} \hline 

{\bf Element}  & \multicolumn{2}{c} {\bf Energy Range} & \multicolumn{1}{c}{\bf Kinetic Energy\tablenotemark{1}} & \multicolumn{1}{c}{\bf Number of} &  \multicolumn{3}{c} {\bf $\quad$ Differential Intensity}\\

 & \multicolumn{2}{c} {\bf (GeV/amu) } & \multicolumn{1}{c}{\bf \^{E} (GeV/amu)} & \multicolumn{1}{c}{\bf Events } & \multicolumn{3}{c} {\bf $\quad$ (m$^2$ s  sr GeV/amu)$^{-1}$}\\ \hline

\\ 
\bf{S    (Z = 16)} & $\quad$      0.8 &        1.0 &        0.9 $\qquad \quad$ &       156 $\quad$ &   $\qquad$(7.5$\;$ &$\pm$ 0.6)&$\;10^{-2}$   \\
 &  $\quad$      1.0 &        1.3 &        1.1 $\qquad \quad$ &       158 $\quad$ &  $\qquad$(6.3$\;$ &$\pm$ 0.5)&$\;10^{-2}$   \\
 &  $\quad$       1.3 &        1.8 &        1.5 $\qquad \quad$ &       170 $\quad$ &  $\qquad$(3.3$\;$ &$\pm$ 0.3)&$\;10^{-2}$   \\
 &  $\quad$      1.8 &        2.3 &        2.0 $\qquad \quad$ &       106 $\quad$ &  $\qquad$(2.0$\;$ &$\pm$ 0.2)&$\;10^{-2}$   \\
 &  $\quad$       12 &         41 &         21 $\qquad \quad$ &      1212 $\quad$ &  $\qquad$(2.1$\;$ &$\pm$ 0.1)&$\;10^{-4}$   \\
 &  $\quad$       41 &        130 &         70 $\qquad \quad$ &       212 $\quad$ &  $\qquad$(1.1$\;$ &$\pm$ 0.1)&$\;10^{-5}$   \\
 &  $\quad$      130 &        412 &        223 $\qquad \quad$ &        32 $\quad$ &  $\qquad$(5.2$\;$ &$\pm$ 0.9)&$\;10^{-7}$   \\
 &  $\quad$      529 &        792 &        644 $\qquad \quad$ &         4 $\quad$ &  $\qquad$(2.1$\;$ & $^{+ 1.6} _{- 1.0}$)&$\;10^{-8}$ \\ 
 &  $\quad$      792 &  &       1257 $\qquad \quad$ &         3 $\quad$ &  $\qquad$(2.6$\;$ & $^{+ 2.5} _{- 1.4}$)&$\;10^{-9}$ \\ 
\vspace{0.05cm} \\ 
 \bf{Ar   (Z = 18)} & $\quad$      0.8 &        1.0 &        0.9 $\qquad \quad$ &        70 $\quad$ & $\qquad$(4.0$\;$ &$\pm$ 0.5)&$\;10^{-2}$   \\
 &  $\quad$      1.0 &        1.3 &        1.1 $\qquad \quad$ &        61 $\quad$ & $\qquad$(2.9$\;$ &$\pm$ 0.4)&$\;10^{-2}$   \\
 &  $\quad$      1.3 &        1.8 &        1.5 $\qquad \quad$ &        66 $\quad$ & $\qquad$(1.5$\;$ &$\pm$ 0.2)&$\;10^{-2}$   \\
 &  $\quad$      1.8 &        2.3 &        2.0 $\qquad \quad$ &        39 $\quad$ & $\qquad$(8.9$\;$ &$\pm$ 1.4)&$\;10^{-3}$   \\
 &  $\quad$       12 &         41 &         21 $\qquad \quad$ &       454 $\quad$ & $\qquad$(7.1$\;$ &$\pm$ 0.3)&$\;10^{-5}$   \\
 &  $\quad$      41 &        130 &         70 $\qquad \quad$ &        85 $\quad$ & $\qquad$(3.8$\;$ &$\pm$ 0.4)&$\;10^{-6}$   \\
 &  $\quad$      130 &        412 &        223 $\qquad \quad$ &        11 $\quad$ & $\qquad$(1.5$\;$ &$\pm$ 0.5)&$\;10^{-7}$   \\
 &  $\quad$      452 &        696 &        558 $\qquad \quad$ &         2 $\quad$ & $\qquad$(1.9$\;$ & $^{+ 2.5} _{- 1.3}$)&$\;10^{-8}$ \\  
 &  $\quad$      696 &  &       1105 $\qquad \quad$ &         1 $\quad$ & $\qquad$(3.2$\;$ & $^{+ 7.3} _{- 2.6}$)&$\;10^{-9}$ \\ 
\vspace{0.05cm} \\ 
 \bf{Ca   (Z = 20)} & $\quad$      0.8 &        1.0 &        0.9 $\qquad \quad$ &        99 $\quad$ & $\qquad$(5.5 $\;$ &$\pm$ 0.6)&$\;10^{-2}$   \\
 &  $\quad$       1.0 &        1.3 &        1.1 $\qquad \quad$ &        79 $\quad$ & $\qquad$(3.7 $\;$ &$\pm$ 0.4)&$\;10^{-2}$   \\
 &  $\quad$      1.3 &        1.8 &        1.5 $\qquad \quad$ &        99 $\quad$ & $\qquad$(2.2 $\;$ &$\pm$ 0.2)&$\;10^{-2}$   \\
 &  $\quad$      1.8 &        2.3 &        2.0 $\qquad \quad$ &        61 $\quad$ & $\qquad$(1.3 $\;$ &$\pm$ 0.2)&$\;10^{-2}$   \\
 &  $\quad$       12 &         41 &         21 $\qquad \quad$ &       610 $\quad$ & $\qquad$(9.9 $\;$ &$\pm$ 0.4)&$\;10^{-5}$   \\
 &  $\quad$       41 &        130 &         70 $\qquad \quad$ &        88 $\quad$ & $\qquad$(4.1 $\;$ &$\pm$ 0.4)&$\;10^{-6}$   \\
 &  $\quad$      130 &        412 &        223 $\qquad \quad$ &        19 $\quad$ & $\qquad$(2.8 $\;$ &$\pm$ 0.6)&$\;10^{-7}$   \\
 &  $\quad$      452 &        696 &        558 $\qquad \quad$ &         3 $\quad$ & $\qquad$(1.8 $\;$ & $^{+ 1.7} _{- 1.0}$)&$\;10^{-8}$ \\ 
 &  $\quad$      696 &  &       1105 $\qquad \quad$ &         1 $\quad$ & $\qquad$(1.9 $\;$ & $^{+ 4.3} _{- 1.6}$)&$\;10^{-9}$ \\ 
\vspace{0.05cm} \\ 
 \bf{Fe   (Z = 26)} &       0.8 &        1.0 &        0.9 $\qquad \quad$ &      1821 $\quad$ & $\qquad$(2.21 $\;$ &$\pm$ 0.05)&$\;10^{-1}$   \\
 &   $\quad$     1.0 &        1.3 &        1.1 $\qquad \quad$ &      1669 $\quad$ & $\qquad$(1.65 $\;$ &$\pm$ 0.04)&$\;10^{-1}$   \\
 &   $\quad$     1.3 &        1.8 &        1.5 $\qquad \quad$ &      2166 $\quad$ & $\qquad$(1.02 $\;$ &$\pm$ 0.02)&$\;10^{-1}$   \\
 &   $\quad$     1.8 &        2.3 &        2.0 $\qquad \quad$ &      1388 $\quad$ & $\qquad$(6.2 $\;$ &$\pm$ 0.2)&$\;10^{-2}$   \\
 &   $\quad$      12 &         23 &         16 $\qquad \quad$ &     10589 $\quad$ & $\qquad$(1.22 $\;$ &$\pm$ 0.02)&$\;10^{-3}$   \\
 &   $\quad$      23 &         41 &         30 $\qquad \quad$ &      4468 $\quad$ & $\qquad$(2.80 $\;$ &$\pm$ 0.04)&$\;10^{-4}$   \\
 &   $\quad$      41 &         92 &         60 $\qquad \quad$ &      2132 $\quad$ & $\qquad$(4.7 $\;$ &$\pm$ 0.1)&$\;10^{-5}$   \\
 &   $\quad$      92 &        412 &        183 $\qquad \quad$ &       543 $\quad$ & $\qquad$(1.9 $\;$ &$\pm$ 0.1)&$\;10^{-6}$   \\
 &   $\quad$     529 &        606 &        566 $\qquad \quad$ &         9 $\quad$ & $\qquad$(1.3 $\;$ & $^{+ 0.6} _{- 0.4}$)&$\;10^{-7}$ \\ 
 &   $\quad$     606 &        895 &        733 $\qquad \quad$ &        12 $\quad$ & $\qquad$(5.2 $\;$ &$\pm$ 1.5)&$\;10^{-8}$   \\
 &   $\quad$     895 &  &       1421 $\qquad \quad$ &        11 $\quad$ & $\qquad$(7.4 $\;$ &$\pm$ 2.2)&$\;10^{-9}$   \\

\end{tabular}
\tablenotetext{1}{see text for definition of $\hat{E}$}
\end{minipage}
\end{table}

\normalsize

\clearpage


\begin{thebibliography}{}
\bibitem[Ahn et al.(2007)]{ahn07} Ahn, H.S. et al., 2007, Proc. of 30th International Cosmic Ray Conference (ICRC)
\bibitem[Agostinelli et al.(2003)]{agostinelli03} Agostintelli, S. et al. 2003, NIM A, 1, 250
\bibitem[Aharonian et al.(2007a)]{aharonian07a} Aharonian, F. et al. 2007a, \prd, 75, 4
\bibitem[Aharonian et al.(2004)]{aharonian04} Aharonian, F. et al. 2004, Nature, 432, 75
\bibitem[Antoni et al.(2005)]{antoni05} Antoni, T. et al. 2005, APh., 24, 1
\bibitem[Bell(1978)]{bell78} Bell, A.R 1978, \mnras, 182, 147
\bibitem[Berezhko \& V\"{o}lk(2006)]{berezhko06} Berezhko, E.G. \& V\"{o}lk, H.J. 2006, \aap, 451, 981
\bibitem[Beuville et al.(1990)]{beuville90} Beuville, E. et al. 1990, NIM A, 288, 157
\bibitem[Cronin et al.(1997)]{cronin97} Cronin, J., Gaisser, T.K., \& Swordy, S.P. 1977, Scientific American, 276, 44 
\bibitem[LaGage \& Cesarsky(1983)]{lagage83} Lagage, P.O. \& Cesarsky, C.J. 1983, \aap, 125, 249L
\bibitem[Derbina et al.(2005)]{derbina05} Derbina, V. et al. 2005, \apjl, 628, L41
\bibitem[Engelmann et al.(1990)]{engelmann90} Engelmann, J. et al. 1990, \aap, 233, 96
\bibitem[Gahbauer et al.(2004)]{gahbauer04} Gahbauer, F., Hermann, G., H\"{o}randel, J., M\"{u}ller D., \& Radu, A.A. 2004, \apj, 607, 333
\bibitem[Gahbauer et al.(2003)]{gahbauer03b} Gahbauer, F., M\"{u}ller D., Hermann, G., H\"{o}randel, J., \& Radu, A. 2003, Proc. 27th ICRC, 1, 2245
\bibitem[Garcia Mu\~{n}oz et al.(1975)]{gmunoz75} Garcia Mu\~{n}oz, M, Mason, G.M, \& Simpson, J.A, 1975, \apjl, 201, L141
\bibitem[Heckmann et al.(1978)]{heckmann78} Heckmann, H.H., Greiner, D.E, Lindstrom, P.J., \& Shwe, H. 1978, \prc, 17, 1735
\bibitem[Juliusson et al.(1972)]{juliusson72} Juliusson, E., Meyer, P., \& M\"{u}ller D., 1972, \prl, 29, 445
\bibitem[L'Heureux et al.(1990)]{lheureux90} L'Heureux, J., Grunsfeld, J.M., Meyer, P., M\"{u}ller, D., \& Swordy, S.P.  1990, NIM A, 295, 246
\bibitem[Mewaldt et al.(2001)]{mewaldt01} Mewaldt, R.A. et al. 2001, Space Science Reviews, 99, 27 
\bibitem[M\"{u}ller et al.(1991)]{muller91} M\"{u}ller, D., Swordy, S.P., Meyer, P., L'Heureux, J., \& Grunsfeld, J. 1991, \apj, 374, 356
\bibitem[Navarra et al.(2003)]{navarra03} Navarra, G. et al. 2003, Proc. 28th International Cosmic Ray Conference, 1, 147
\bibitem[Panov et al.(2006)]{panov06} Panov, A. et al. 2006, ArXiv e-print, astro-ph/0612377
\bibitem[Romero-Wolf(2005)]{romero05b} Romero-Wolf, A. 2005, MSc. Thesis, The University of Chicago, Unpublished
\bibitem[Seo et al.(2006)]{seo06} Seo, E.S. et al. 2006, Advances in Space Sciences, in press
\bibitem[Westfall et al.(1979)]{westfall79} Westfall, G. D., Wilson, L. W., Lindstrom, P. J., Crawford, H. J., Greiner, D. E., \& Heckman, H. H., \prc, 19, 1309
\bibitem[Wiedenbeck et al.(1999)]{wiedenbeck99} Wiedenbeck, M.E et al. 1999, \apjl, 523, L61
\bibitem[Yanasak et al.(2001)]{yanasak01} Yanasak, N. E. et al., 1975, \apj, 563, 768
\bibitem[Zei et al.(2007)]{zei07} Zei, R. et al., 2007, Proc. of 30th ICRC
\end{thebibliography}
\end{document}